\documentclass[fleqn,usenatbib]{mnras}

\usepackage{newtxtext,newtxmath}

\usepackage[T1]{fontenc}

\DeclareRobustCommand{\VAN}[3]{#2}
\let\VANthebibliography\thebibliography
\def\thebibliography{\DeclareRobustCommand{\VAN}[3]{##3}\VANthebibliography}

\usepackage{graphicx}	
\usepackage{amsmath}	
\usepackage[anythingbreaks]{breakurl}

\graphicspath{{./fig/}}
\usepackage{bm}

\newcommand{\EQ}{\begin{equation}}
\newcommand{\EN}{\end{equation}}
\newcommand{\EQA}{\begin{eqnarray}}
\newcommand{\ENA}{\end{eqnarray}}

\newcommand{\Eq}[1]{Eq.~(\ref{#1})}

\newcommand{\App}[1]{Appendix~\ref{#1}}

\newcommand{\Sec}[1]{Sect.~\ref{#1}}

\newcommand{\Fig}[1]{Fig.~\ref{#1}}
\newcommand{\FFig}[1]{Figure~\ref{#1}}

\newcommand{\Figs}[2]{Figs.~\ref{#1} and \ref{#2}}
\newcommand{\Figss}[2]{Figs.~\ref{#1}--\ref{#2}}
\newcommand{\Tab}[1]{Table~\ref{#1}}

\newcommand{\bra}[1]{\langle #1\rangle}
\newcommand{\bbra}[1]{\left\langle #1\right\rangle}

{}
{}
{}

{}
{}
{}
{}
{}
{}
{}
{}
{}
{}
{}
{}
{}
{}
{}
{}
{}

{}

{}

{}
{}
{}

\newcommand{\eee}{\hat{\mbox{\boldmath $e$}} {}}

\newcommand{\kk}{\bm{k}}

\newcommand{\xx}{\bm{x}}

\newcommand{\BB}{\bm{B}}

\newcommand{\JJ}{\bm{J}}

\newcommand{\AAA}{\bm{A}}

\newcommand{\uu}{\bm{u}}

\newcommand{\ff}{\mbox{\boldmath $f$} {}}

\newcommand{\nab}{{\bm{\nabla}}}

\newcommand{\SSSS}{\mbox{\boldmath ${\sf S}$} {}}

\newcommand{\ii}{{\rm i}}

\newcommand{\DD}{{\rm D} {}}

\newcommand{\dd}{{\rm d} {}}

\newcommand{\const}{{\rm const}  {}}

\def\degr{\hbox{$^\circ$}}

\def\Sp{\mbox{\rm Sp}}
\def\Ma{\mbox{\rm Ma}}

\def\Pm{\mbox{\rm Pr}_{\rm M}}
\def\Rm{\mbox{\rm Re}_{\rm M}}

\def\Rey{\mbox{\rm Re}}

\def\EK{E_{\rm K}}
\def\EM{E_{\rm M}}

\def\csz{c_{\rm s0}}

\def\cp{c_{\rm p}}
\def\cv{c_{\rm v}}

\def\cs{c_{\rm s}}

\def\xiM{\xi_{\rm M}}

\def\kf{k_{\rm f}}

\def\EM{E_{\rm M}}

\def\kin{{\rm kin}}
\def\sat{{\rm sat}}

\def\epsK{\epsilon_{\rm K}}

\def\vArms{v_{\rm A}^{\rm rms}}

\def\Brms{B_{\rm rms}}

\def\urms{u_{\rm rms}}

\newcommand{\SKEW}{{\rm skew}}
\newcommand{\KURT}{{\rm kurt}}

\newcommand{\cm}{\,{\rm cm}}

\newcommand{\kms}{\,{\rm km\,s}^{-1}}

\newcommand{\kpc}{\,{\rm kpc}}

\newcommand{\RM}{{\rm RM}}

\def\apj{ApJ}
\def\apjl{ApJL}

\def\prd{PhRvD}
\def\pre{PhRvE}
\def\prf{PhRvF}
\def\prl{PhRvL}

\def\araa{ARA\&A}

\def\aap{A\&A}
\def\mnras{MNRAS}

\title[Kazantsev spectra in the small-scale dynamo]{Batchelor, Saffman, and Kazantsev spectra in galactic small-scale dynamos}

\author[A. Brandenburg et al.]{
Axel Brandenburg$^{1,2,3,4}$\thanks{E-mail:brandenb@nordita.org},
Hongzhe Zhou$^{1,5}$, and
Ramkishor Sharma$^{1,2}$
\\
$^1$Nordita, KTH Royal Institute of Technology and Stockholm University, Hannes Alfv\'ens v\"ag 12, SE-10691 Stockholm, Sweden \\
$^2$The Oskar Klein Centre, Department of Astronomy, Stockholm University, AlbaNova, SE-10691 Stockholm, Sweden\\
$^3$McWilliams Center for Cosmology and Department of Physics, Carnegie Mellon University, 5000 Forbes Ave, Pittsburgh, PA 15213, USA\\
$^4$School of Natural Sciences and Medicine, Ilia State University, 3-5 Cholokashvili Avenue, 0194 Tbilisi, Georgia\\
$^5$Tsung-Dao Lee Institute, Shanghai Jiao Tong University, 800 Dongchuan Road, Shanghai 200240, People's Republic of China
}

\date{\today}

\pubyear{2022}

\begin{document}
\label{firstpage}
\pagerange{\pageref{firstpage}--\pageref{lastpage}}
\maketitle

\begin{abstract}
The magnetic fields in galaxy clusters and probably also in the
interstellar medium are believed to be generated by a small-scale dynamo.
Theoretically, during its kinematic stage, it is characterized by a
Kazantsev spectrum, which peaks at the resistive scale.
It is only slightly shallower than the Saffman spectrum that is expected
for random and causally connected magnetic fields.
Causally disconnected fields have the even steeper Batchelor spectrum.
Here we show that all three spectra are present in the small-scale dynamo.
During the kinematic stage, the Batchelor spectrum occurs on scales
larger than the energy-carrying scale of the turbulence, and the Kazantsev
spectrum on smaller scales within the inertial range of the turbulence --
even for a magnetic Prandtl number of unity.
In the saturated state, the dynamo develops a Saffman spectrum on
large scales, suggestive of the build-up of long-range correlations.
At large magnetic Prandtl numbers, elongated structures are seen
in synthetic synchrotron emission maps showing the parity-even $E$
polarization.
We also observe a significant excess in the $E$ polarization over the
parity-odd $B$ polarization at subresistive scales, and a deficiency at
larger scales.
This finding is at odds with the observed excess in the Galactic microwave
foreground emission, which is believed to be associated with larger scales.
The $E$ and $B$ polarizations may be highly non-Gaussian and skewed in the kinematic regime of the dynamo.
For dust emission, however, the polarized emission is always nearly
Gaussian, and the excess in the $E$ polarization is much weaker.
\end{abstract}

\begin{keywords}
dynamo -- MHD -- polarization -- turbulence -- galaxies: magnetic fields
\end{keywords}

\section{Introduction}

The possibility of small-scale magnetohydrodynamic (MHD) dynamos has
been studied since the early work of \cite{Bat50}, who assumed that
the statistical properties of the magnetic field agree with those of
vorticity.
However, the nowadays accepted theory of small-scale dynamos was developed
only later by \cite{Kaz68}.
However, the topic of small-scale dynamos moved somewhat into the
background with the discovery of large-scale dynamos driven by the
helicity or $\alpha$ effect \citep{SKR66, Mof78, KR80}.
With the advent of direct numerical simulations (DNS) of turbulence,
the study of small-scale dynamos was picked up again by \cite{MFP81} and
\cite{Kida+91}.
In both studies, the addition of kinetic helicity had only a minor
effect on the result, which was due to too small domain sizes.
Early convection-driven dynamos with rotation \citep{MP89, Nor+92}
therefore also essentially counted under this category.
The work of \cite{Kaz68} became routinely quoted only since the 2000s
when simulations began to reproduce what is nowadays often referred to
as the Kazantsev spectrum \citep{Scheko04, HBD04}.

The original theory of \cite{Kaz68} was linear, so it only described the
early kinematic growth phase of the dynamo.
Furthermore, it assumed that the velocity was smooth and of large
scale only.
In the framework of turbulence, this could be realized if the magnetic
Prandtl number, $\Pm\equiv\nu/\eta$, is large, i.e., if the viscosity
$\nu$ is much larger than the magnetic diffusivity $\eta$, making then
the velocity field much smoother than the magnetic field.
This situation is applicable to galaxies and galaxy clusters, but not
to stars and other denser bodies.

Kazantsev's theory yielded as the eigenfunction a magnetic energy spectrum
$\EM(k)$ proportional to $k^{3/2}K_0(k/k_\eta)$, where $k$ is the
wavenumber, $K_0$ is the
Macdonald function of order zero or the modified Bessel function of the
second kind, and $k_\eta=(4\gamma/15\eta)^{1/2}$ is the resistive cutoff
wavenumber \citep{Kulsrud+Anderson92} with $\gamma$ being the growth rate.
Such scaling was indeed confirmed in a number of different DNS
\citep{Scheko04, HBD04, BRS22}.
However, it is important to recall that this scaling is only expected
for large values of $\Pm$.
In the opposite limit of $\Pm\ll1$, the spectral slope may be smaller.
\cite{Bra+18} confirmed a $k^{7/6}$ scaling for $\Pm=0.1$, as was
previously discussed by \cite{SB14}.

In the meantime, there has been a significant amount of work on decaying
turbulence.
Much of this was motivated by applications to the early universe
\citep{BEO96, CHB01, BJ04}.
An important question here is how rapidly the magnetic energy decays
and how rapidly the correlation length of the turbulence increases.
It has been argued that this may depend on the slope of the subinertial
range spectrum, i.e., on the exponent $\alpha$ in the magnetic energy
spectrum $\EM(k)\propto k^\alpha$.
Here, the subinertial range is the low wavenumber part of the spectrum
below the peak wavenumber \citep{Olesen97}.
Above the peak, we usually have the inertial range, where the velocity
is expected to have a Kolmogorov $k^{-5/3}$ spectrum, which is followed
by the viscous subrange above some viscous cutoff wavenumber.

\cite{Olesen97} found the possibility of inverse cascading, i.e.,
a temporal increase of the spectral power for small $k$ and a rapid
decrease of the correlation length $\xiM$ when $\alpha$ is large enough.
The inverse correlation length $\xiM^{-1}$ is usually close to the
position of the peak of the spectrum of $\EM(k)$.
In forced turbulence, we have $\xiM^{-1}\approx\kf$, where $\kf$
is the forcing wavenumber, but in decaying turbulence, its value is
time-dependent (and decreasing).

In the simulations of \cite{CHB01}, an initial $k^4$ spectrum was assumed.
The value $\alpha=4$ was argued to be a general consequence of the
requirement of causality in the early Universe, i.e., the requirement
that the magnetic field $\BB$ is uncorrelated over different positions,
and the fact that $\nab\cdot\BB=0$ \citep{DC03}.
A $k^4$ subinertial range spectrum is usually referred to as a
Batchelor spectrum.
DNS have shown that, in the presence of magnetic helicity,
a $k^4$ spectrum develops automatically, even when the initial spectrum
was shallower, e.g., $\propto k^2$, which is called a Saffman spectrum
in hydrodynamic turbulence without helicity \citep{Saff67}.
However, subsequent work showed that this is only true because of the
presence of magnetic helicity and that, then, non-helical turbulence
with an initial Saffman spectrum preserves its initial $k^2$ slope
\citep{2017PhRvE..96e3105R,Bran+17}.

Many of the MHD decay studies where done for magnetically dominated
turbulence \citep{BKT15, BK17}, i.e., the initial magnetic energy density
is large compared with the kinetic energy density of the turbulence.
This precludes the investigation of dynamo action, i.e., the conversion
of kinetic energy into magnetic.
Simulations of \cite{Bra+19} showed that a nearly exponential increase
of magnetic energy is still possible for some period of time when the
initial magnetic energy density is small enough.

To summarize, in decaying MHD turbulence, the magnetic energy spectrum
can have a $k^2$ or a $k^4$ spectrum
\cite[see the discussion in Sect.~3.5 of][]{kandu2019},
depending on the initial conditions.
For nonhelical turbulence, in particular, there is no reason to expect
a $k^4$ spectrum, unless the causality argument of \cite{DC03} can
be invoked.
Earlier work did show a $k^4$ spectrum of the magnetic field in the
kinetically dominated case; see Fig.~8 of \cite{Bra+19}, but this was
in the presence of helicity.
Moreover, there was no indication of a Kazantsev $k^{3/2}$ spectrum.
This could perhaps be related to the fact that in those simulations,
the magnetic Prandtl number was chosen to be unity, i.e., not $\gg1$.
There remained therefore the question, how the Batchelor $k^4$ spectrum,
the Saffman $k^2$ spectrum, and the Kazantsev $k^{3/2}$ spectrum are
related to each other.
Disentangling this is the motivating topic of this paper.

We consider forced turbulence with a weak initial seed magnetic field.
We consider DNS with an isothermal equation of state using
a resolution of $N^3=1024^3$ mesh points, which is
still not large enough to cover all turbulent subranges in one simulation,
but experimenting with selected subranges remains affordable.
We therefore compare simulations with different values of $\Pm$ and $\kf$.
It will turn out that all three spectra, the Batchelor, Saffman, and
Kazantsev spectra are being realized in the small-scale dynamo problem
if the range of available wavenumbers is large enough; the Batchelor
and Saffman spectra are being found in the subinertial range during the
kinematic and saturated growth phases, respectively, and the Kazantsev
$k^{3/2}$ spectrum is found in what corresponds to the magnetic inertial
range during the kinematic phase.
In the saturated stage, however, it changes to a declining spectrum,
which is typically close to the Kolmogorov $k^{-5/3}$ spectrum, or the
Iroshnikov-Kraichnan $k^{-3/2}$ slope, whose theoretical foundation
is still subject to research \citep{Bol05, Bol06, Schekochihin2020}.

To make contact with observations, it is essential to determine diagnostic
quantities.
At our disposal are observations of synchrotron and dust emission, both
causing linear polarization.
Linear polarization is expressed in terms of Stokes $Q$ and $U$
parameters, but those are not independent of the rotation of the frame.
This problem is well known in cosmology and, since the pioneering work
of \cite{SZ97} and \cite{Kamionkowski+97} it is therefore customary
to transform $Q$ and $U$ into its rotationally invariant components,
called $E$ and $B$.
These names are supposed to remind the reader of gradient-like and
curl-like fields, but have otherwise nothing to do with electric or
magnetic fields.
However, the cosmological interpretation of $E$ and $B$ is strongly
affected by Galactic foreground emission from dust and synchrotron
emission, depending on the wavelength \citep{Choi+Page15}.
This has led to an interest in studies of the $E$ and $B$ polarization
from MHD waves and turbulence \citep{Caldwell} with applications to
emission from the interstellar medium \citep[ISM; see][]{Kri+18,
Bracco+19, Bra19}, galaxies \citep{BB20b, BF20}, and even the Sun
\citep{Bra+19b, Bra19EB, Bra20IAUS, Prabhu+20, Prabhu+21}.
It was initially thought that the parity-odd polarization can be
used as a proxy of magnetic helicity, but this is only true in systems
that are inhomogeneous along the line of sight \citep{Bra+19b}.
Observationally, we also know that in the ISM, the $E$ polarization
exceeds the $B$ polarization by a factor of about two \citep{PlanckXXX}.
The reason behind this is not entirely clear, and we are still learning
from the diversity of results that have been accumulated in recent years
for different systems.
This is the reason why we analyze $E$ and $B$ also for the present
simulations.
For the present small-scale dynamo simulations, it turns out, however,
that the results for $E$ and $B$ are not very sensitive to the properties
of the hydrodynamic flow, and that even a non-isothermal, two-phase
flows can reproduce similar $E$ and $B$ patterns.
One such result will be presented at the very end.
We begin by presenting the equations for our isothermal setup that will
be used in the main part of the paper.
Next, we compare the energy spectra for five different dynamo runs,
before presenting their $E$ and $B$ signatures.

\section{Our model}

\subsection{Basic equations}

We consider weakly compressible turbulence with an isothermal equation
of state and constant speed of sound $\cs$, where the pressure $p$
is proportional to the density $\rho$, i.e., $p=\rho\cs^2$.
We solve for the magnetic vector potential $\AAA$, so the magnetic
field is $\BB=\nab\times\AAA$.
The full set of evolution equations for $\AAA$, the velocity $\uu$,
and the logarithmic density $\ln\rho$ is given by
\begin{equation}
\frac{\partial\AAA}{\partial t}=\uu\times\BB+\eta\nabla^2\AAA,
\end{equation}
\begin{equation}
\frac{\DD\uu}{\DD t}=\ff-\cs^2\nab\ln\rho+
\frac{1}{\rho}\left[\JJ\times\BB+\nab\cdot(2\rho\nu\SSSS)\right],
\label{DuDt}
\end{equation}
\begin{equation}
\frac{\DD\ln\rho}{\DD t}=-\nab\cdot\uu,
\end{equation}
where $\JJ=\nab\times\BB/\mu_0$ is the current density and
$\mu_0$ is the vacuum permeability, ${\sf S}_{ij}=
(\partial_i u_j+\partial_j u_i)/2-\delta_{ij}\nab\cdot\uu/3$
are the components of the rate-of-strain tensor $\SSSS$,
and $\ff$ is a nonhelical forcing function consisting
of plane waves with wavevector $\kk$.
It is proportional to $(\eee\times\kk)\,e^{\ii\kk\cdot\xx}$, where $\xx$
is position and $\eee$ is a randomly chosen unit vector that is not
aligned with $\kk$.
The wavevector $\kk$ changes randomly at each time step, making the
forcing function therefore $\delta$ correlated in time.
We select the wavevectors $\kk$ randomly from a finite set of vectors
whose components are multiples of $k_1\equiv2\pi/L$, where $L$ is the
side length of our Cartesian domain of volume $L^3$.
This forcing function has been used in many earlier papers
\citep[e.g.][]{HBD04}.

\subsection{Governing parameters and diagnostics}
\label{Governing}

For all our simulations, we use the {\sc Pencil Code} \citep{JOSS}, which
is an explicit code whose time step is given by the Courant-Friedrich-Levy
condition and therefore scales inversely with the maximum wave speed.
We arrange the forcing strength such that the Mach number based on the
rms velocity of the turbulence, $\Ma=\urms/\cs$, is around 0.1.
This choice ensures that the turbulence is sufficiently subsonic and
therefore close to the incompressible limit, but not so small that the
sound speed, which is the main factor limiting the time step, does not
exceed $\urms$ by an unreasonably large amount.

Our governing parameters are the fluid and magnetic Reynolds numbers,
defined here as
\begin{equation}
\Rey=\urms/\nu\kf,\quad
\Rm=\urms/\eta\kf,
\end{equation}
respectively.
Thus, $\Pm=\Rm/\Rey$.
We usually try to keep these two Reynolds numbers as large as possible.
As a rule of thumb, one may say that the product of
$(\kf/k_1)\times\max(\Rey,\Rm)$ should not exceed the mesh size
by a large factor.
Usually, the simulation would ``crash'', i.e., the turbulent energy
cannot be dissipated anymore at the highest wavenumbers.
Even if the simulation does not crash, the accuracy of the results may
be affected.
However, since potential artifacts are expected to affect mostly the
high wavenumber part of the spectrum, we might still trust the low
wavenumber part.
It should be kept in mind that a small Reynolds number also causes
artifacts, because the simulation becomes too diffusive, so it is
essential to choose just the right value.
This can only be decided in the context of and through the comparison
with simulations for other parameters.

In any dynamo problem, an important output parameter is the growth rate
\begin{equation}
\gamma=\bra{\dd\ln\Brms/\dd t}_{\rm kin},
\end{equation}
where the subscript `kin' denotes a time average over the kinematic stage.
We normalize $\gamma$ by the turnover rate and denote it by a tilde,
i.e., $\tilde{\gamma}=\gamma/\urms\kf$, where $\urms$ is taken from the
kinematic phase of the dynamo.

We define kinetic and magnetic energy spectra that are
normalized such that $\int\EK(k)\,\dd k=\bra{\uu^2}/2$
and $\int\EM(k)\,\dd k=\bra{\BB^2}/2\mu_0\rho_0$, respectively,
where $\rho_0$ is the mean density.
Here, angle brackets without subscript denote volume averages.
Note that the integrals over our energy spectra have units of energy
per unit mass.
We always present time-averaged spectra, which is straightforward
for the kinetic energy because they are statistically stationary,
and it is therefore also straightforward for the magnetic energy in the
saturated regime, but in the kinematic phase, $\EM(k,t)$ is exponentially
growing with the rate $2\gamma$, so we average the compensated spectra,
$\bra{e^{-2\gamma t}\EM(k,t)}_{\rm kin}$, over a suitable time interval
where the product is stationary; see also \cite{SB14} where this was done.

When plotting spectra, we normalize $k$ by the viscous dissipation
wavenumber
\begin{equation}
k_\nu=(\epsK/\nu^3)^{1/4},
\end{equation}
where $\epsK=\bra{2\rho\nu\SSSS^2}$ is the kinetic energy dissipation rate.
Furthermore, one would often present compensated spectra by normalizing
them with $k^{5/3}\epsK^{-2/3}$, which would not only make it nondimensional,
but this would then also yield a flat inertial range, whose mean value,
$\bra{\EK(k)\,k^{5/3}\epsK^{-2/3}}_{\rm inert}\equiv C_{\rm Kol}$ would
yield the Kolmogorov constant, $C_{\rm Kol}$, and the subscript `inert'
denotes averaging over the inertial subrange.
Here, we are not so much interested in the inertial range and would
instead like the original slopes to be persevered.
Therefore, we normalize with a $k$-independent, fixed value
$\kf^{5/3}\epsK^{-2/3}$, which would still allow us to read off the
approximate Kolmogorov constant at the position of the peak of the
spectrum normalized in this way.

We select forcing wavevectors $\kk$ from a narrow band of vectors with
$\kf-\delta k/2\leq |\kk| < \kf+\delta k/2$, where $\delta k$ is chosen
such that the number of possible wavevectors does not exceed 10,000.
For large values of $\kf/k_1$, we therefore reduce $\delta k$.
This manifests itself in the kinetic energy of spectra, which then have
a progressively sharper spike as the forcing wavenumber is increased.

\subsection{Polarized synchrotron and dust emissions}

In the ISM, one can measure the magnetic field through linearly polarized
synchrotron and dust emission.
The Stokes $Q$ and $U$ parameters can be combined into a complex
polarization $P=Q+\ii U$, which is given by a line-of-sight integration
\citep{Pac70},
\begin{equation}
P=-\int_0^\infty \epsilon({\cal B})\,e^{2\ii\phi(z)\lambda^2}\dd z,
\end{equation}
where ${\cal B}=B_x+\ii B_y$ is the complex magnetic field
in the observational plane, the emissivity is approximated as
$\epsilon({\cal B})=\epsilon_0 {\cal B}^2$ for synchrotron emission and
$\epsilon({\cal B})=\epsilon_0 {\cal B}^2/|{\cal B}|^2$ for dust emission
\citep{PlanckXX, Bracco+19}, and
$\phi(z)=-K\int_0^z n_{\rm e} B_z\, \dd z'$ is the Faraday depth with
$n_{\rm e}$ being the density of thermal electrons and $K$ is a constant.
The prefactor on the emissivity, $\epsilon_0$, is here taken as a
constant, whose actual value is not important as we present our results
only in normalized forms.
The Faraday depth across a slab of length $L$ is $\phi(L)$ and gives
the rotation measure as ${\rm RM}=\phi(L)/2$ \citep{BS14}.
Observationally, RM can be determined by varying the wavelength
$\lambda$, i.e., ${\rm RM}=\dd\arg(P)/\dd\lambda^2$, where $\arg(P)$
is the phase of $P$.
If the electron density is approximately constant, RM is proportional
to the line-of-sight integrated line-of-sight component of the magnetic
field.
Significant amount of work has been done to establish the relation
between the spectra of the magnetic field and those of $\RM$
\citep{CR09, BS13, SBS18, Seta+22}.

As observational diagnostics, we present the line-of-sight averaged
magnetic field, $\bra{B_z}_z(x,y)$, as a proxy for RM when
$n_e=\const$, and the linear polarization $P(x,y)$ is computed
in the absence of Faraday depolarization, i.e., $\lambda=0$.
In the following, we convert $P(x,y)$ into the parity-even
and parity-odd $E$ and $B$ polarizations by computing \citep{SZ97,
Kamionkowski+97, Bra+19b}
\begin{equation}
R(x,y)={\cal F}^{-1}\left[(k_x-\ii k_y)^2 {\cal F}(P)\right],
\end{equation}
where ${\cal F}(P)=\int P(\xx_\perp)\,e^{\ii\kk_\perp\cdot\xx_\perp}
\,\dd^2\xx_\perp/(2\pi)^2$ is a
function of $\kk_\perp\equiv(k_x,k_y)$ and denotes the Fourier transformation
over the $\xx_\perp\equiv(x,y)$ plane, ${\cal F}^{-1}$ denotes the
inverse transformation, and $R=E+\ii B$.
Thus, the real and imaginary parts of $R(x,y)$ give the 
$E$ and $B$ polarizations, respectively.

\section{Results on the energy spectra}

\subsection{Presentation of the results}

We summarize our simulations in \Tab{Tsum}.
Here, the runs are listed in the order of decreasing $\kf$ and increasing
$\Rey$.
Again, the tildes denote normalized quantities, i.e.,
$\tilde{k}_{\rm f}=\kf/k_1$, $\tilde{k}_\nu=k_\nu/k_1$,
$\tilde{v}_{\rm A}^{\rm rms}=\vArms/\cs$, where
$\vArms=\Brms/\sqrt{\mu_0\rho_0}$ is used to quantify the magnetic field
strength as an Alfv\'en velocity.
The values of $\vArms$ refer to the saturated state,
but all others refer to the kinematic phase.
During saturation, Ma decreases, especially when $\Pm=1$; see
\App{Changes} for details.

\begin{figure}\begin{center}
\includegraphics[width=\columnwidth]{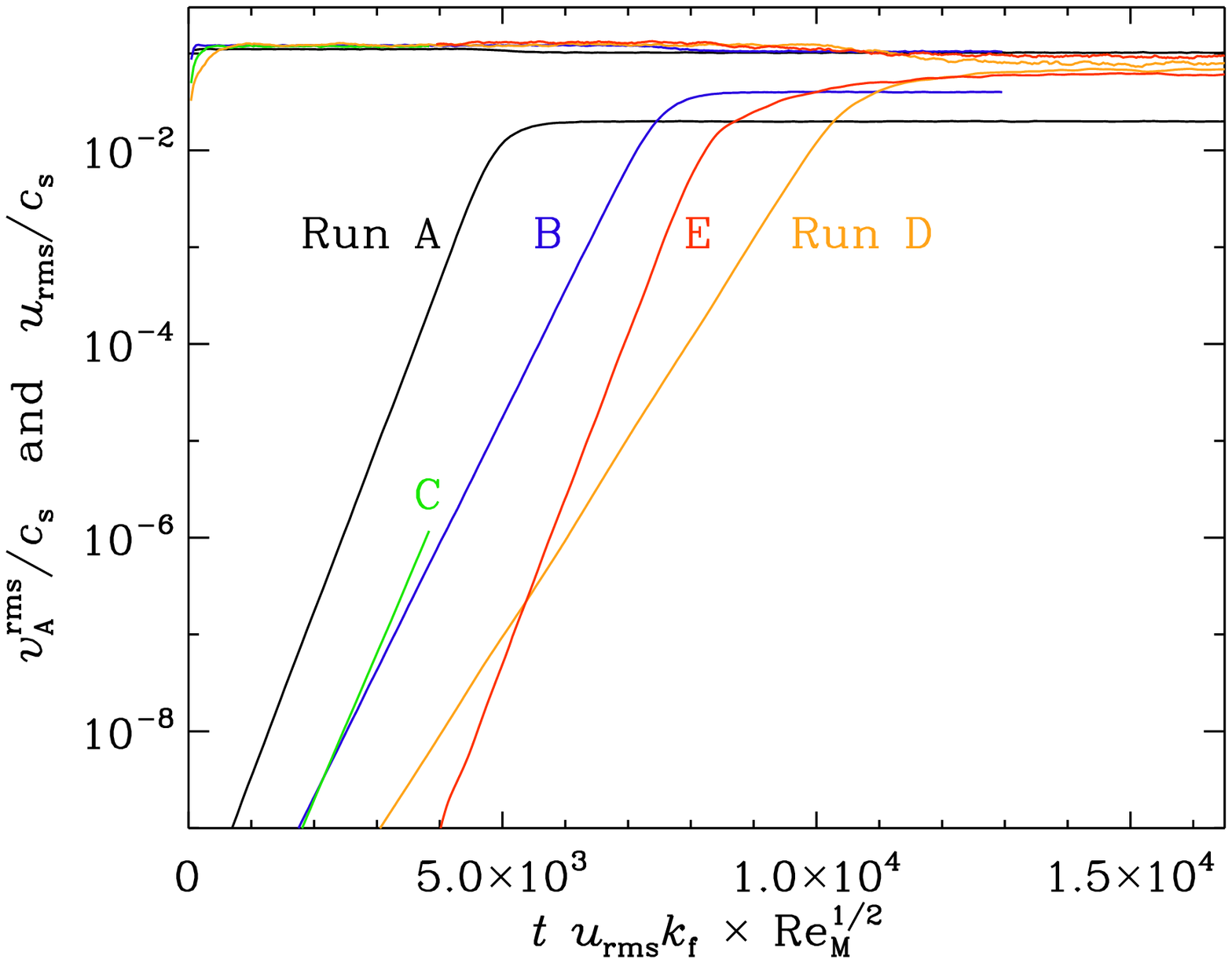}
\end{center}\caption[]{
Time series of $\urms$ and $\vArms=\Brms/\sqrt{\mu_0\rho_0}$,
normalized by $\cs$.
}\label{pcomp_kaz}\end{figure}

We begin by discussing Run~B with $\Pm=1$ and then Runs~C and D
with $\Pm=10$ and 30.
The reason we discuss Run~A with $\kf/k_1=120$ later is because we
first want to motivate the need for going to such an extremely high
value of $\kf$.
Finally, we present with Run~E a case where $\Pm=1$, so as to show that
the choice of large magnetic Prandtl numbers is not necessary
to obtain a Kazantsev spectrum in the kinematic phase.
For all runs, the evolution of $\urms$ and $\Brms$ is shown in \Fig{pcomp_kaz}.
Here, the time axis is normalized by the turnover time, $1/\urms\kf$,
and scaled by the square root of the Reynolds number, so as to have
comparable saturation times.

\begin{table}
\centering
\caption{
Summary of the simulations presented in this paper.
}\label{Tsum}
\begin{tabular}{crrrccrrr} 
\hline
Run & $\Ma$ & $\tilde{k}_{\rm f}$ & $\tilde{k}_\nu$ & $\tilde{\gamma}$ &
$\tilde{v}_{\rm A}^{\rm rms}$ & $\Rey$ & $\Rm$ & $\Pm$ \\
\hline
A & 0.111 &120 & 764 & 0.022 & 0.01 &   31 &   31 &  1 \\ 
B & 0.121 & 30 & 389 & 0.027 & 0.03 &   81 &   81 &  1 \\ 
C & 0.118 & 10 & 106 & 0.085 &  ... &   59 &  590 & 10 \\ 
D & 0.122 &  4 &  62 & 0.130 & 0.08 &  100 & 3000 & 30 \\ 
E & 0.130 &  4 & 461 & 0.158 & 0.10 & 1600 & 1600 &  1 \\ 
\hline
\end{tabular}
\end{table}

\subsection{Subinertial range during saturation}
\label{Subinertial}

\begin{figure}\begin{center}
\includegraphics[width=\columnwidth]{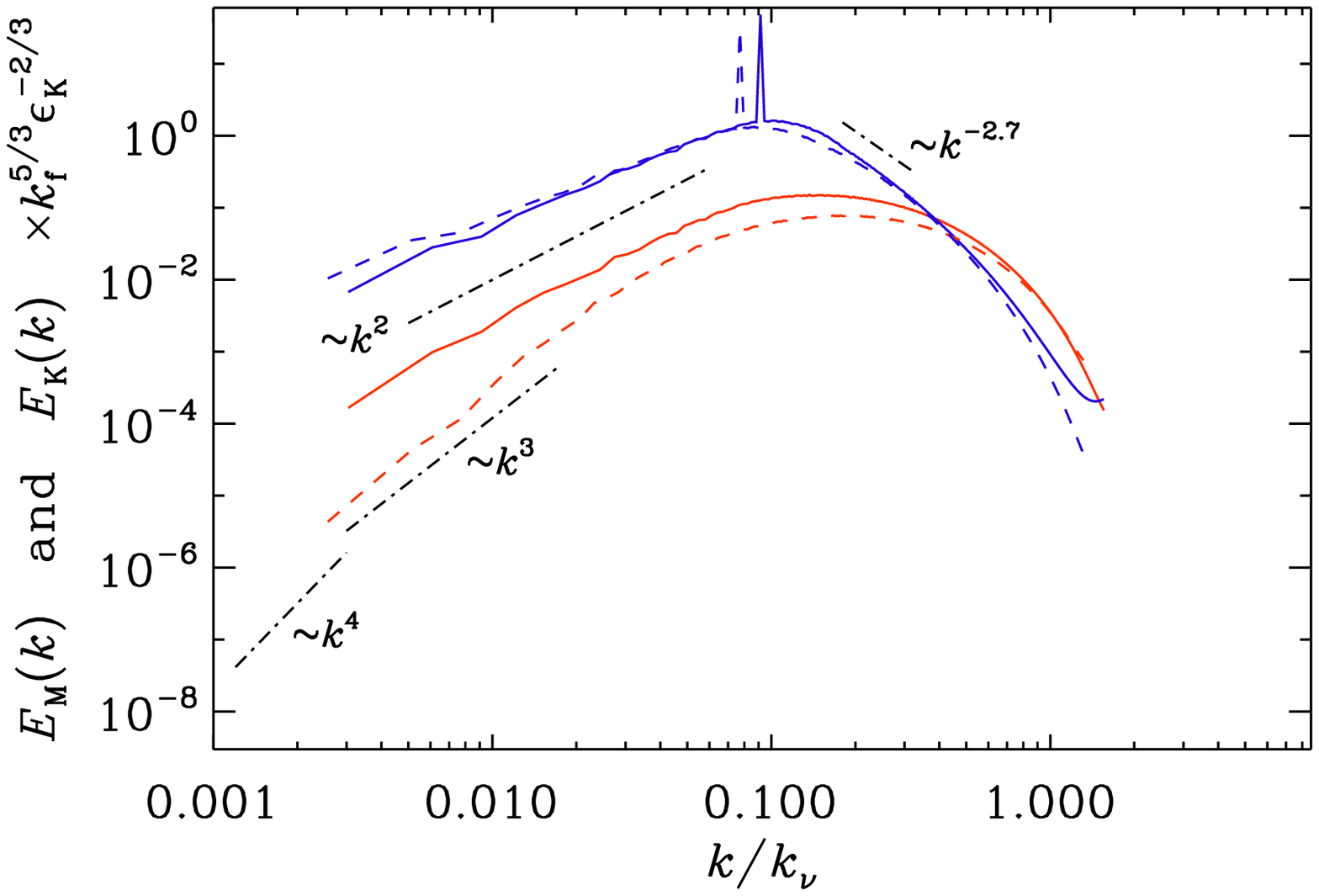}
\end{center}\caption[]{
Kinetic (blue) and magnetic (red) energy spectra for Run~B during
the kinematic (dotted lines) and the saturated (solid lines) stages.
}\label{pspec_comp_kinsat}\end{figure}

To begin with, we consider a case with $\Pm=1$ and $\kf/k_1=30$;
see \Fig{pspec_comp_kinsat}.
For this and the following spectra, we have fixed the ranges on the
abscissa and ordinate so as to facilitate comparison between them.
The position of the peak of the spectrum is clearly visible as a sharp
spike, as explained in \Sec{Governing}.
We see that during the kinematic and saturated phases of the dynamo,
indicated by dashed and solid lines, respectively, the kinetic energy
spectrum always has a clear $k^2$ subinertial range, while the magnetic
field has a steeper subinertial range during the kinematic growth phase
(closer to $k^3$), but becomes proportional to $k^2$ during the saturated
phase.

We note that there is no Kazantsev $k^{3/2}$ slope in the kinematic range.
This may have two reasons: $\Pm$ is not large enough or the turbulent
inertial range is too short.
Therefore, we consider next a run with larger values of $\Pm$.
Later we also reconsider runs with $\Pm=1$ using both a larger and
a smaller value of $\kf$.

\subsection{Emergence of the Kazantsev slope}

\begin{figure}\begin{center}
\includegraphics[width=\columnwidth]{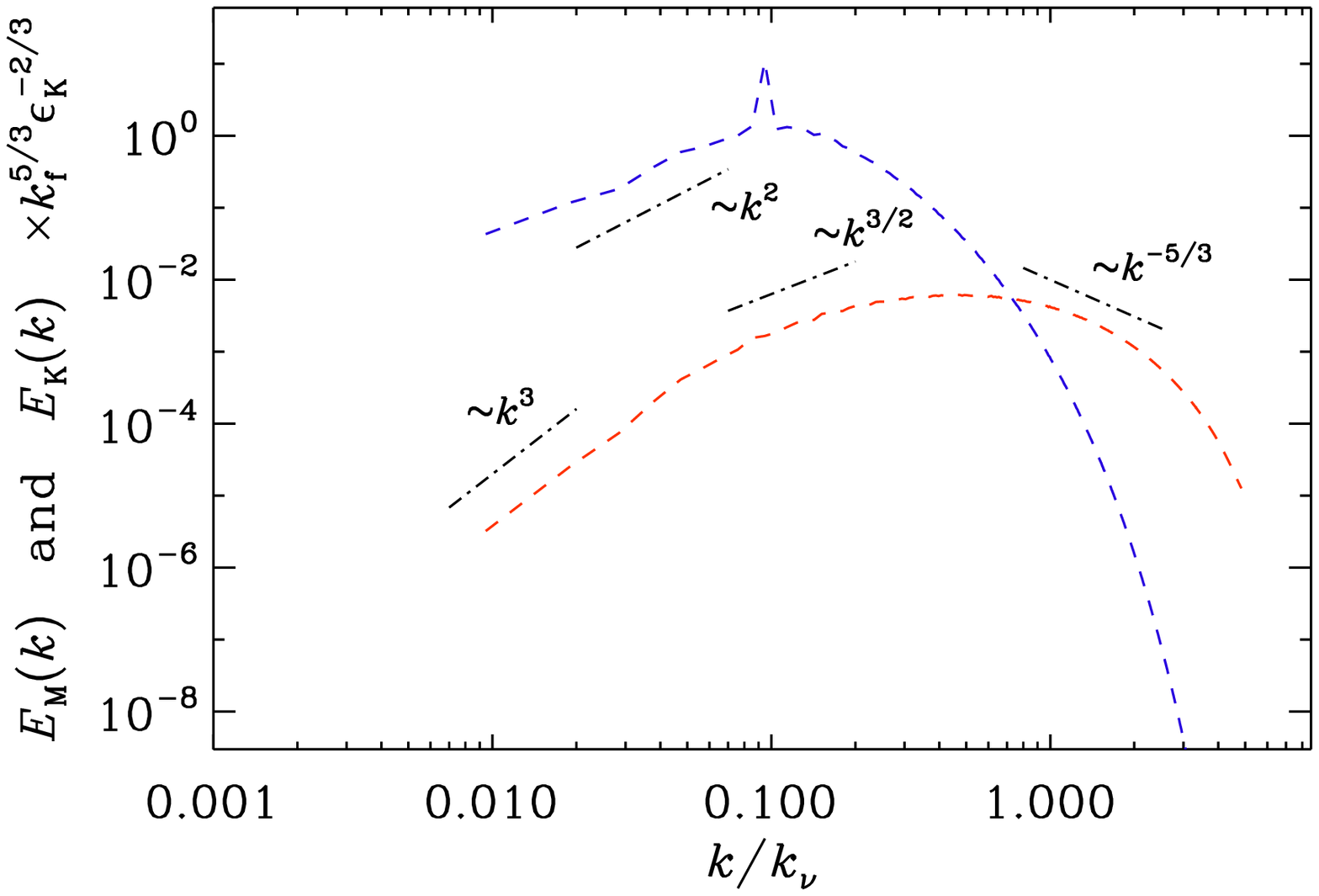}
\end{center}\caption[]{
Similar to \Fig{pspec_comp_kinsat}, but for Run~C,
except that the spectra are only shown for the kinematic phase.
}\label{pspec_comp_kinsat_Pm10}\end{figure}

We now consider Runs~C and D with larger magnetic Prandtl numbers,
$\Pm=10$ and 30, respectively.
Of these two runs, only Run~D has been run into saturation,
because we expect their nonlinear behaviors to be similar.
In these runs, shown in \Figs{pspec_comp_kinsat_Pm10}
{pspec_comp_kinsat_Pm30}, we also decrease the value
of $\kf$ to 10 and 4, respectively.
In both cases, we clearly see the emergence of a Kazantsev $k^{3/2}$
subrange for $k>\kf$.
For Run~D, we also see that the Kazantsev slope disappears in the
saturated state.
We show $k^{-5/3}$ slopes for comparison, but it is clear that
there is insufficient dynamical range to identify a proper
magnetic inertial range.
We still see in the kinematic phase the $k^3$ subrange for $k<\kf$.
However, it is possible that the actual slope of the kinematic subrange
spectrum is steeper, and that we just did not have enough scale separation
between the lowest wavenumber $k_1$ and the forcing wavenumber $\kf$.
Therefore, we now consider a more extreme case with even more scale
separation.

\begin{figure}\begin{center}
\includegraphics[width=\columnwidth]{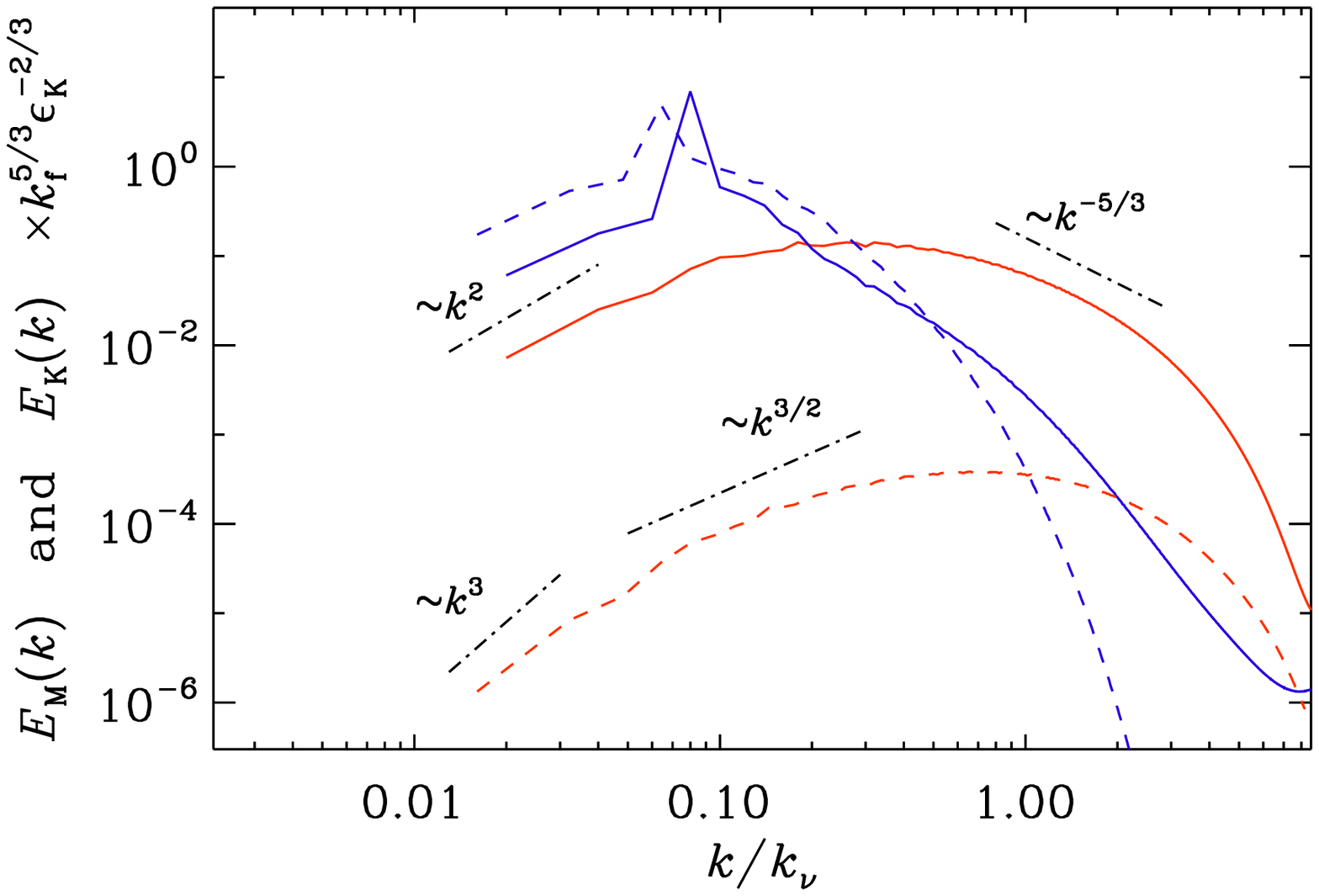}
\end{center}\caption[]{
Similar to \Fig{pspec_comp_kinsat}, but for Run~D.
}\label{pspec_comp_kinsat_Pm30}\end{figure}

\subsection{Batchelor spectrum in the kinematic stage}

\begin{figure}\begin{center}
\includegraphics[width=\columnwidth]{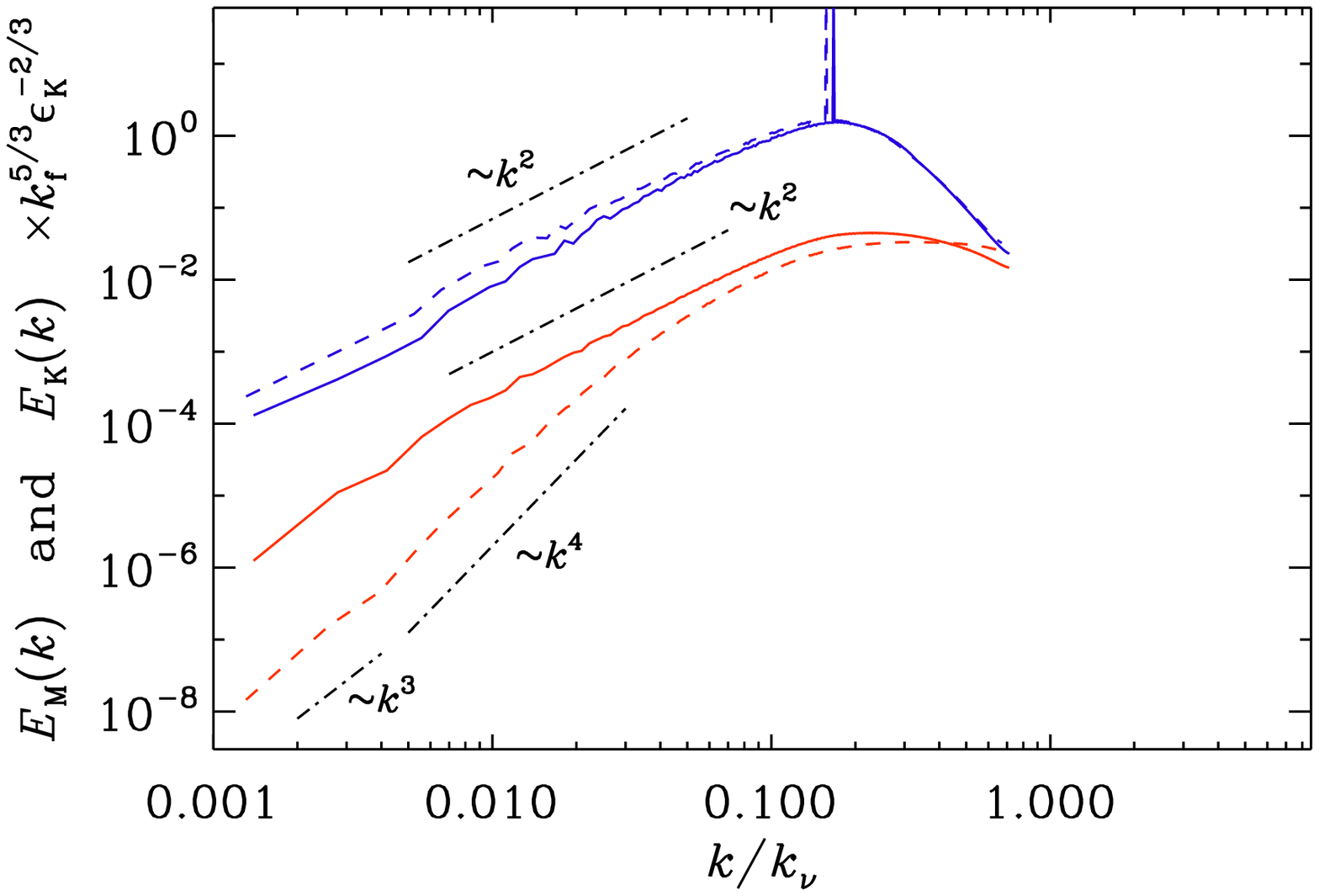}
\end{center}\caption[]{
Similar to \Fig{pspec_comp_kinsat}, but for Run~A.
Note the $k^4$ subrange of $\EM(k)$ in the kinematic stage,
but also evidence for slight contamination at very small $k$.
}\label{pspec_comp_kinsat_Pm1}\end{figure}

To see whether the $k^3$ subrange slope found in \Sec{Subinertial} was
a consequence of still insufficient scale separation, we now consider a
more extreme case with a four times larger value, namely $\kf/k_1=120$;
see \Fig{pspec_comp_kinsat_Pm1}.
We now see that there is indeed a $k^4$ Batchelor subinertial range.
Interestingly, we also see that near the very lowest wavenumbers in the
domain, the spectrum does become slightly shallower.
This suggests that the spectrum at those low wavenumbers,
$1\leq k/k_1\leq 3$, is indeed contaminated by finite size effects
of the computational domain.

\begin{figure}\begin{center}
\includegraphics[width=\columnwidth]{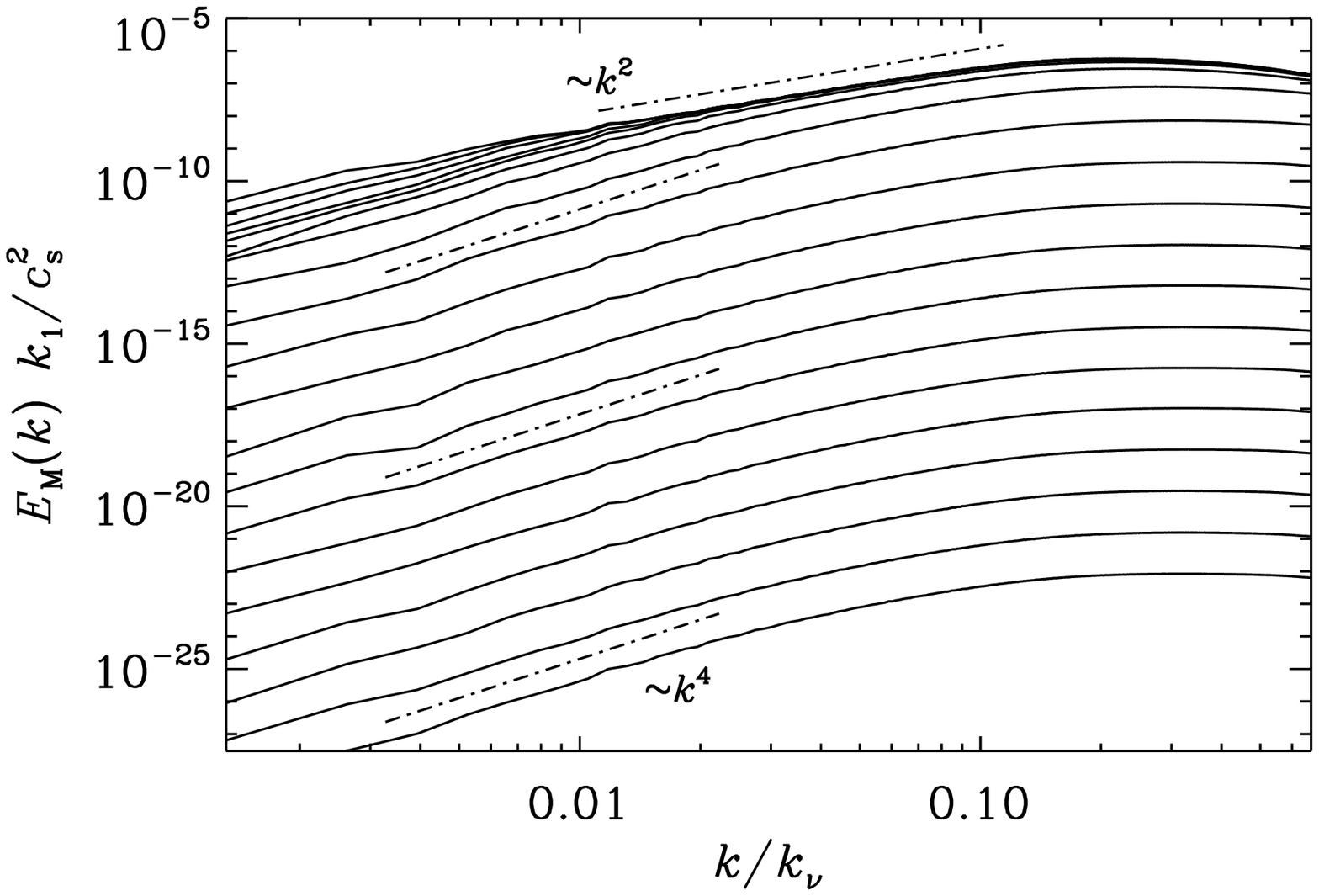}
\end{center}\caption[]{
Unnormalized spectra for Run~A showing that the $k^4$ subrange existed
throughout the entire kinematic phase.
}\label{pspec_select}\end{figure}

To demonstrate that the $k^4$ subrange existed throughout the entire
kinematic phase, we show in \Fig{pspec_select} unnormalized spectra
for Run~A in regular time intervals during the kinematic stage and less
frequently during the saturated stage, where the low wavenumber part is
seen to grow slightly.
The final slope during the saturated stages is $\propto k^2$, just like
the kinetic energy spectrum; see \Fig{pspec_comp_kinsat_Pm1}.

The theoretical reason for the $k^4$ spectrum in the kinematic regime
could lie in the statistical independence of different patches in space.
This led \cite{DC03} to suggest that primordial magnetic fields in the
early Universe must always have a $k^4$ spectrum, as was already assumed
in \cite{CHB01}.
The velocity field, by contrast, is driven by the magnetic field in a
causal fashion, and it always shows a $k^2$ spectrum.
When the magnetic field saturates, different patches are again no longer
uncorrelated.
This may explain the transition from a $k^4$ to a $k^2$ spectrum as a
magnetic field saturates.
To see the reason, note that for an isotropic turbulent magnetic field, its energy spectrum
can be expanded at small $k$ as
\begin{equation}
E_\text{M}(k)=\frac{I_Bk^2}{4\pi^2}+\frac{I_{L_\text{M}}k^4}{24\pi^2}+\mathcal{O}(k^6).
\end{equation}
Here,
\begin{equation}
I_B=\int \bbra{\bm B(\bm x)\cdot\bm B(\bm x+\bm r)}\ \text{d}^3r
\end{equation}
and
\begin{equation}
I_{L_\text{M}}=-\int \bbra{\bm B(\bm x)\cdot\bm B(\bm x+\bm r)}r^2\ \text{d}^3r
\end{equation}
are the magnetic Saffman and magnetic Loitsyansky integrals, respectively \citep{Hosking+Schekochihin20}.
Thus a transition from $k^4$ to $k^2$ in the subinertial
range for the magnetic energy spectrum suggests that $I_B$
has grown from zero to a finite value.
This in turn implies that the magnetic field has built
up some long-range correlations during nonlinear saturation,
in the sense that $\bbra{\bm B(\bm x)\cdot\bm B(\bm x+\bm r)}$
can decay as slowly as $r^{-3}$ as $r\to\infty$.
Such long-range correlations might result
from those of the velocity field, the latter being
thoroughly discussed in \cite{Hosking+22}.

\subsection{Kazantsev spectrum at $\Pm=1$}

\begin{figure}\begin{center}
\includegraphics[width=\columnwidth]{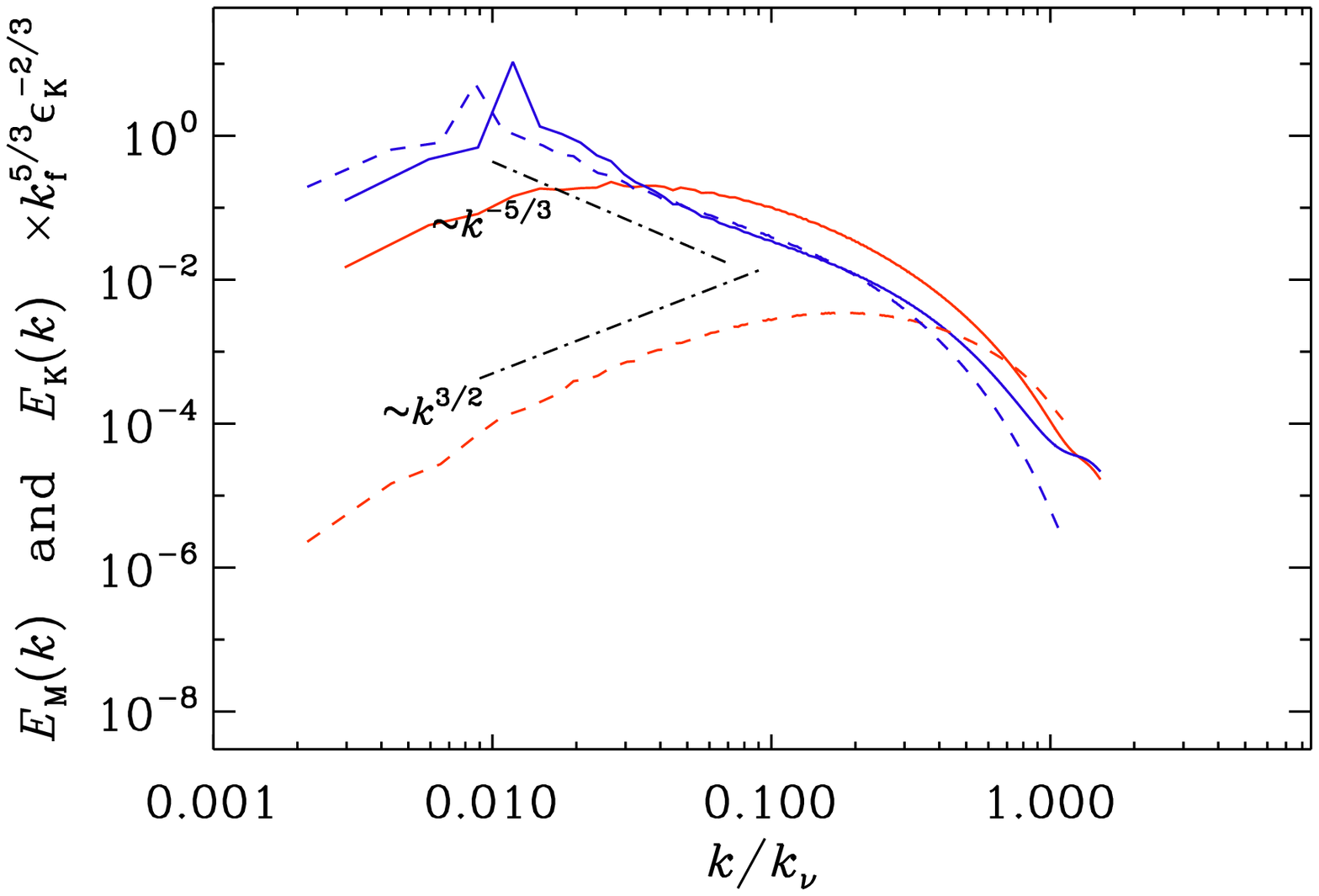}
\end{center}\caption[]{
Similar to \Fig{pspec_comp_kinsat}, but for Run~E.
Note the well developed $k^{3/2}$ Kazantsev slope for $\EM(k)$ during
the kinematic stage in the range where $\EK(k)$ shows a $k^{-5/3}$ 
Kolmogorov subrange with bottleneck in the kinematic stage.
}\label{pspec_comp_kinsat_Pm1_kf4}\end{figure}

We recall that, in order to see the Kazantsev spectrum, we increased
$\Pm$ from one to 10 and 30, but we also decreased $\kf/k_1$ from 30
to 10 and four.
We now show that a larger value of $\Pm$ was helpful in achieving
dynamo action, but it was not essential for obtaining the Kazantsev
spectrum.
The important point is rather that the Kazantsev spectrum is really
a small-scale phenomenon and is not present in the subinertial range.
Therefore, all that is required is a long enough inertial range.
To show this more clearly, we now present a case with $\Pm=1$ and
$\kf/k_1=4$; see \Fig{pspec_comp_kinsat_Pm1_kf4}.
We clearly see the Kazantsev spectrum during the kinematic
stage of the dynamo.

In \Fig{pspec_comp_kinsat_Pm1_kf4}, we also notice that,
in the wavenumber
interval with the Kazantsev slope in the magnetic energy spectrum, we also have
a Kolmogorov inertial range in the kinetic energy spectrum together with
a slight uprise near the dissipation wavenumber.
This uprise is well known in turbulence theory and is referred to as
the bottleneck effect \citep{Falk94}.
It is a phenomenon that is particularly clear in the three-dimensional
spectra presented here, but it is less pronounced in the one-dimensional
spectra considered in observations such as wind tunnel experiment,
which has a simple mathematical reason \citep{Dobler+03}.

\subsection{Inertial-range dynamo action}

\begin{figure*}
\centering
\includegraphics[width=\textwidth]{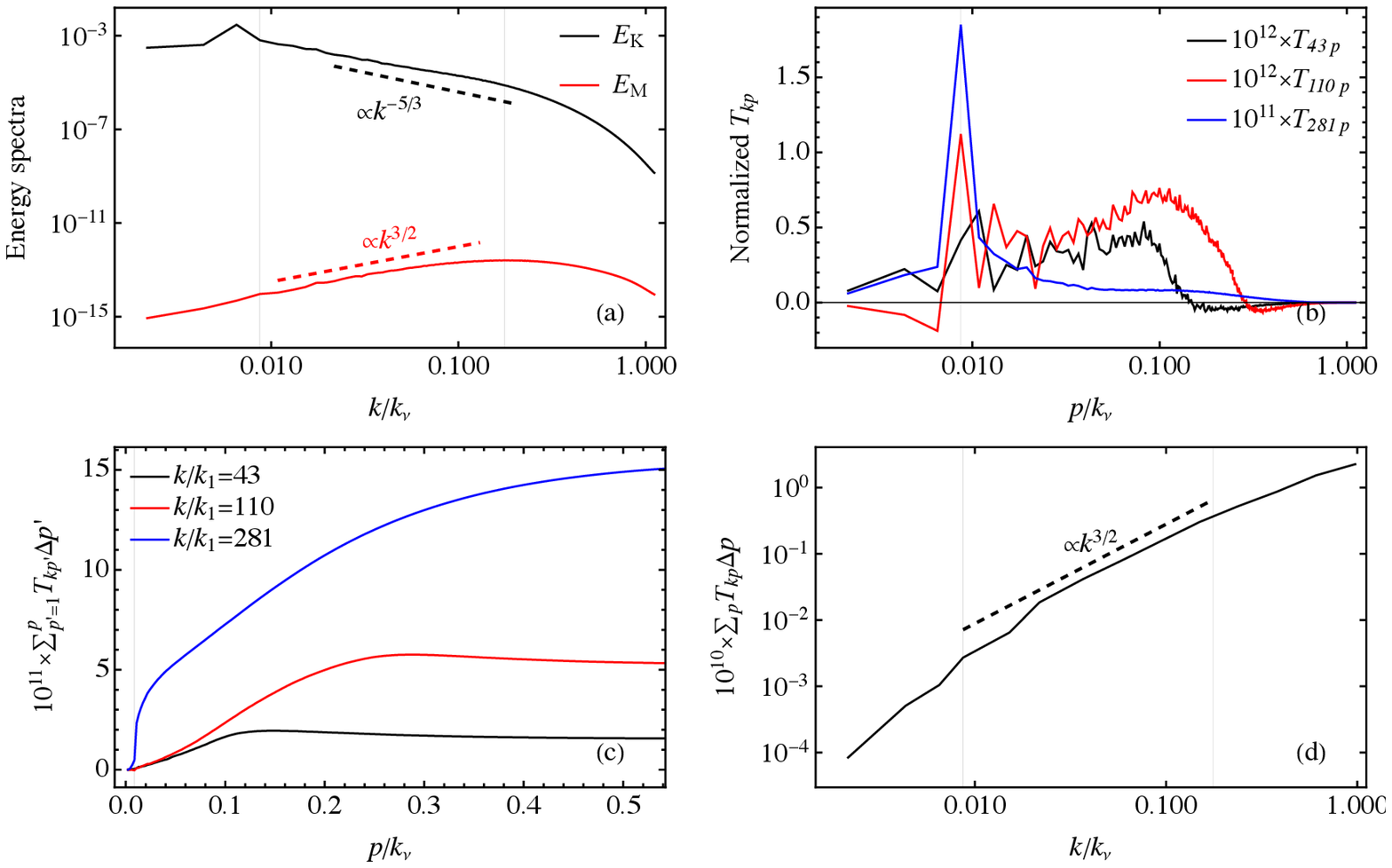}
\caption{Results for run E.
(a) The velocity and magnetic energy spectra at $t=300.4$.
(b) The shell-to-shell energy transfer rate $T_{kp}$,
for three selected values $k=43$, $110$, and $281$.
(c) The accumulative transfer rate.
(d) The net transfer rate.
Data in panels (b), (c), and (d) is obtained by
averaging over $10$ snapshots at $t=300.4, 300.8,\cdots,304.0$.}
\label{fig:hz_tra}
\end{figure*}

Run~E has demonstrated that the Kazantsev spectrum extends well into
the inertial range to length scales above the viscous scale.
Is this truly caused by dynamo action in the inertial range or, perhaps,
a nonlocal artifact caused by the strong spike in kinetic energy at the
driving scale?
As explained above, those peaks can be very large for large values of
$\kf/k_1$, although for Run~E, this spike is no longer so pronounced.
To analyze the spectral energy flux, we follow the procedure of
\cite{BKT15} and compute the spectral transfers
\begin{equation}
T_{kpq}=\bra{\JJ_k\cdot(\uu_p\times\BB_q)}\quad\mbox{and}\quad
T_{kp}=\bra{\JJ_k\cdot(\uu_p\times\BB)}.
\end{equation}
Here, the subscripts on the vectors indicate linearly spaced wavenumbers
of filtering over concentric shells in wavenumber space.
In \Fig{fig:hz_tra}, we present the result for the kinematic stage.
\FFig{fig:hz_tra}(a) shows the energy spectra for velocity and magnetic
fields, and the vertical lines mark the forcing scale at
$\kf/k_1=4$ (left), and the peak of the magnetic energy spectrum
at $k_\text{peak}/k_1\simeq 81$ (right).
In Fig.~\ref{fig:hz_tra}(b), we show the shell-to-shell transfer
rates $T_{kp}$ at $k/k_1=43$, $110$, and $281$, which corresponds
to $k/k_\nu=0.09$, $0.24$, and $0.61$.
We see that there are strong peaks at the forcing scale,
but also considerable contributions from smaller scales.
Note that \Fig{fig:hz_tra}(b) is obtained by averaging $T_{kp}$
for $10$ snapshots, taken at intervals $\simeq0.17/\urms\kf$.
For $k$ near or below $k_\text{peak}$, the contribution
at $p=\kf$ fluctuates, and sometimes becomes negative.
This is because the velocity field at $\kf$ is mostly
driven by the random-in-time forcing, which rapidly
changes its direction.
At large $k$, near $k_\nu$, the velocity at that scale has become
too small to drive dynamo action, and the energy transfer into the
magnetic field mostly comes from the energy-dominant eddy at $\kf$
tangling magnetic field lines.
This explains the persistent and dominant peak
for the blue curve in \Fig{fig:hz_tra}(b).

To identify where the dominant contribution comes from, we compute the accumulative
transfer rate, $\int_0^p T_{kp'}\text{d}p'$; see Fig.~\ref{fig:hz_tra}(c).
It is clear that the velocity modes in the inertial range
contribute roughly equally to a given magnetic mode.
Although the flow velocity has a strong peak near the forcing scale,
it does not play a particularly significant role in the kinematic dynamo phase.
In panel (d) we show the net transfer rate into shell $k$,
$\int T_{kp}\text{d}p$, which scales as $k^{3/2}$ in the inertial range.
Given that the magnetic energy spectrum is also $\propto k^{3/2}$ in the
inertial range, this suggests that the dynamo growth rate $\text{d}\ln
E_\text{M}/\text{d}t$ is independent of the wavenumber $k$, as would
have been expected from the Kazantsev model, although, strictly speaking,
the model is expected to be valid only for $\Pm\gg1$.

\section{Diagnostics of different subranges}

\subsection{Diagnostic images and spectra}
\label{Diagnostics}

\begin{figure}\begin{center}
\includegraphics[width=\columnwidth]{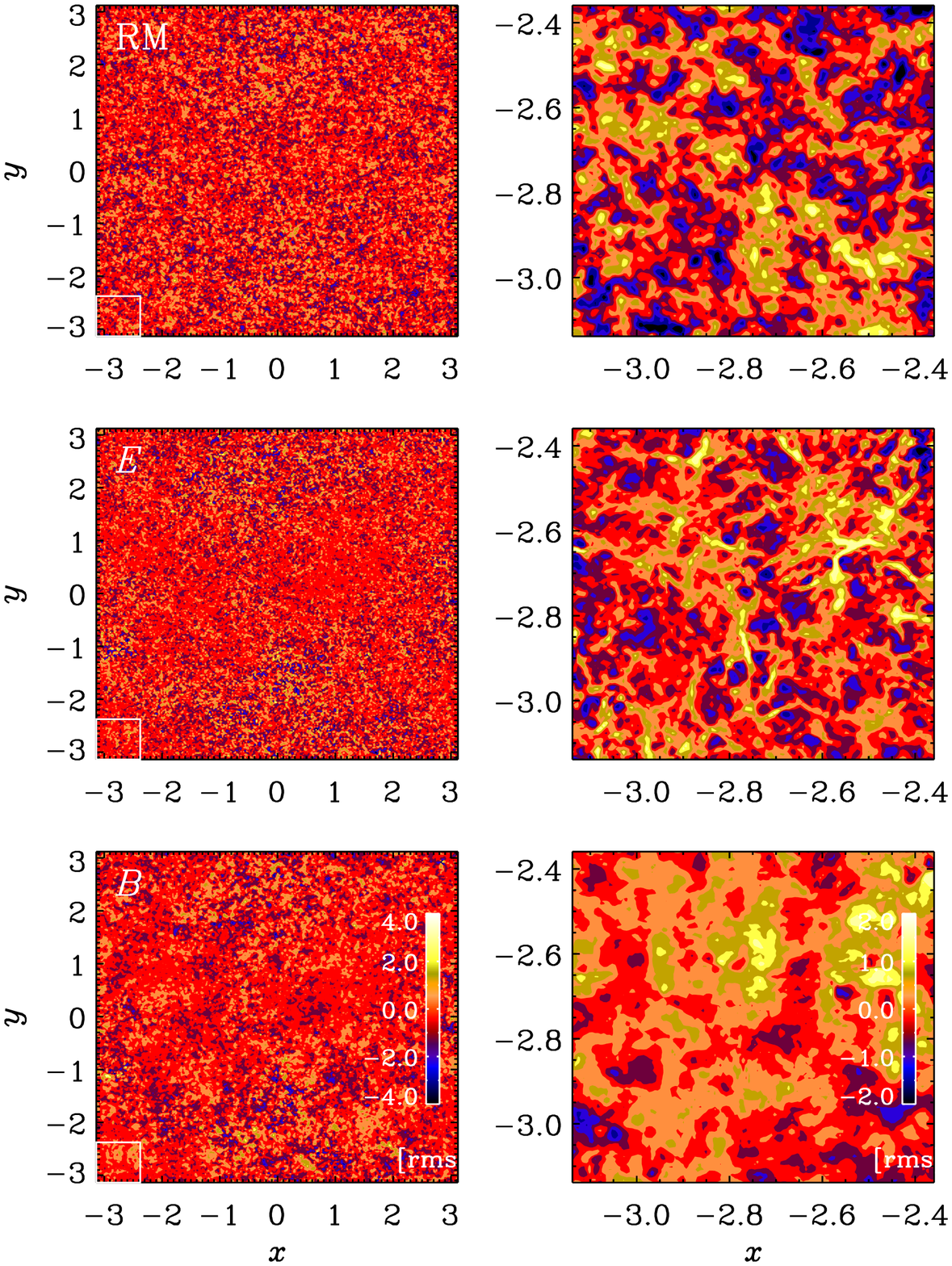}
\end{center}\caption[]{
Diagnostic slices of RM (top row),
$E$ (middle row), and $B$ (bottom row), for Run~E during the kinematic stage.
The small white squares on the left column mark the part
that is shown enlarged on the right column.
All quantities are normalized by their rms value and the color bars
for the enlarged frames are clipped at $\pm2$ times the rms value,
while those for the full frames are clipped at $\pm4$ times the rms value.
}\label{pdiag_planes}\end{figure}

In \Fig{pdiag_planes}, we present for Run~E synthetic radio images of
$E(x,y)$, $B(x,y)$, and
$\bra{B_z}_z(x,y)$, where the latter will simply be denoted by ${\rm RM}$.
The structures are rather small, so we also show an enlarged presentation
of $1/8^2$ of the image, as indicated by the white box on the corresponding
full images.

\begin{figure}\begin{center}
\includegraphics[width=\columnwidth]{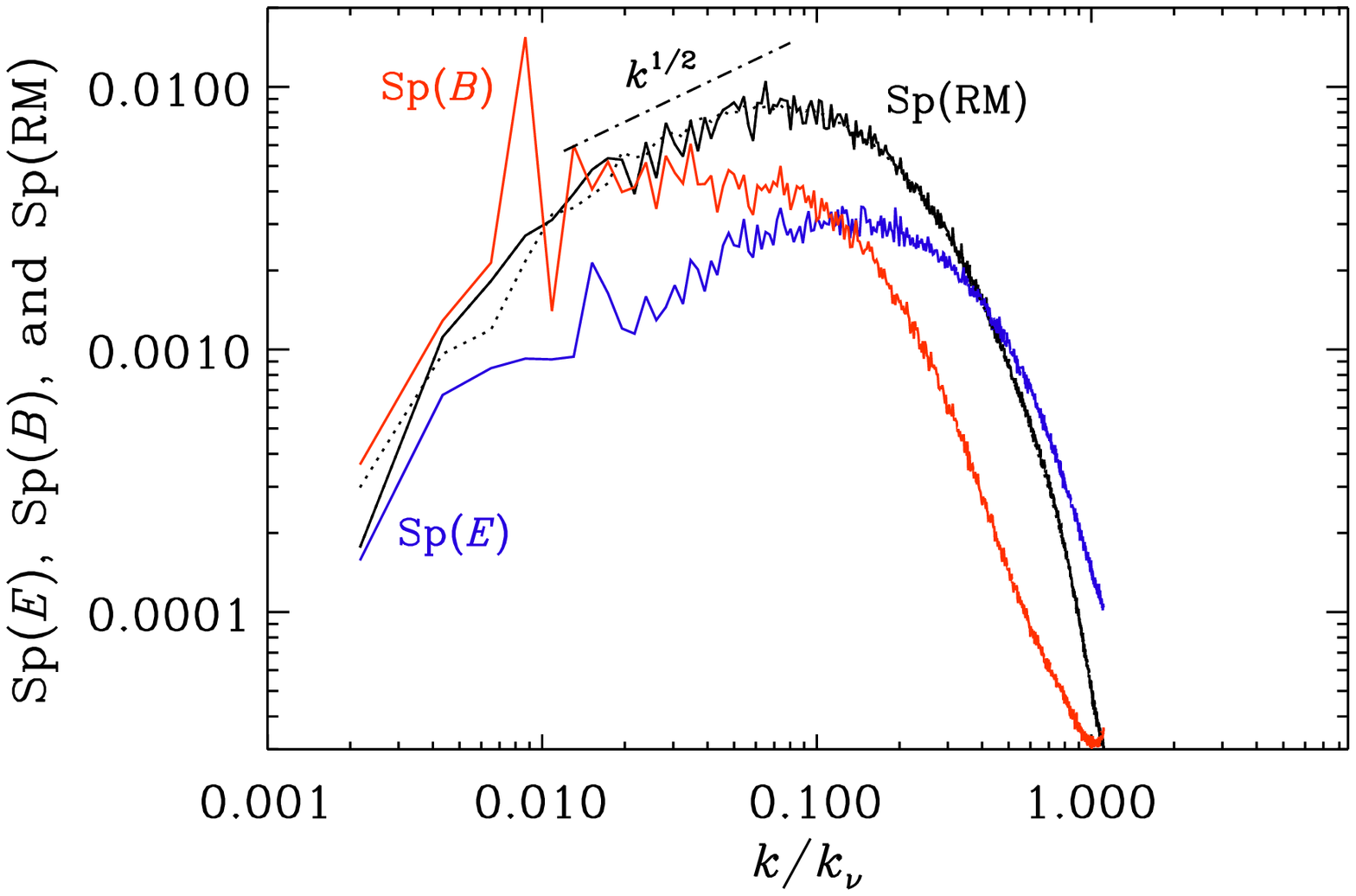}
\end{center}\caption[]{
Diagnostic spectra $\Sp(E)$ (blue line), $\Sp(B)$ (red line), and
$\Sp(\RM)$ (black line) for Run~E during the kinematic stage.
The dotted line gives $\EM(k)/k$, normalized so that it nearly
overlaps with $\Sp(\RM)$.
}\label{pdiag_spec}\end{figure}

\begin{figure}\begin{center}
\includegraphics[width=\columnwidth]{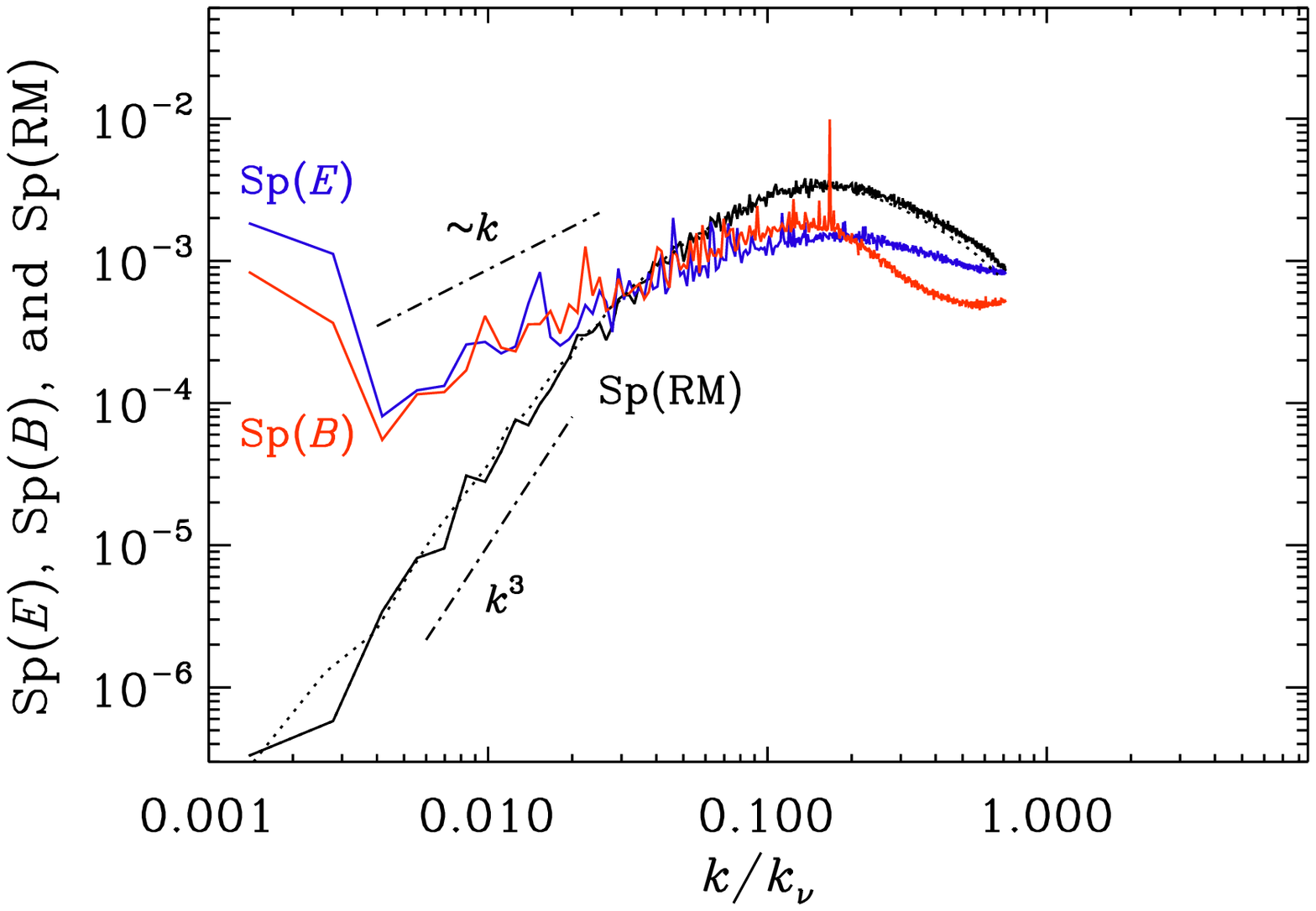}
\end{center}\caption[]{
Same as \Fig{pdiag_spec}, but for Run~A during the kinematic stage.
}\label{pdiag_spec_D1024_Pm1e_kf120}\end{figure}

Next, we present diagnostic spectra from our two-dimensional
synchrotron images for Run~E during the kinematic stage.
They are denoted by $\Sp(E)$, $\Sp(B)$, and $\Sp(\RM)$ and
are normalized such that $\int\Sp(\RM)\,\dd k=1$ and
$\int[\Sp(E)+\Sp(B)]\,\dd k=1$; see \Fig{pdiag_spec}.
For comparison, we also overplot $\EM(k)/k$, suitably normalized,
which is seen to agree with $\Sp(\RM)$.
It turns out that in $0.01\leq k/k_\nu\leq0.1$,
$\Sp(\RM)$ and $\Sp(E)$ are proportional to $k^{1/2}$.
On the other hand, $\Sp(B)$ is flat; see \Tab{Tscl} for the approximate
wavenumber scalings of $\EM(k)$, $\Sp(\RM)$, $\Sp(E)$, and $\Sp(B)$
for the inertial range of Run~E.

For white noise in two dimensions, we would expect a linearly increasing
spectrum.
In the present case, this is indeed the case for the $E$ and
$B$ polarizations in the subinertial range of Run~A; see \Tab{Tadd}
and \Fig{pdiag_spec_D1024_Pm1e_kf120}.

\begin{table}
\centering
\caption{
Approximate wavenumber scalings of $\EM(k)$, $\Sp(\RM)$, $\Sp(E)$,
and $\Sp(B)$ for the subinertial range of Run~A and the inertial
range of Run~E during the kinematic phase.
}\label{Tscl}
\begin{tabular}{clccrrcc} 
\hline
Run & range & $\EM(k)$ & $\Sp(\RM)$ & $\Sp(E)$ & $\Sp(B)$ \\
\hline
A & subinertial & $k^4$ & $k^3$ & $k^1$ & $k^1$ \\
E &    inertial & $k^{3/2}$ & $k^{1/2}$ & $k^{1/2}$ & $k^0$ \\
\hline
\end{tabular}
\end{table}

The correspondence between the exponent $\alpha$ in $\EM(k)\propto k^\alpha$
and $\alpha_{\RM}$ in $\Sp(\RM)\propto k^{\alpha_{\RM}}$ with
$\alpha_{\RM}=\alpha-1$ is explained by the line-of-sight integration,
which removes the spatial dependence in one direction.
For Run~E, this is also seen for $\alpha_E$ in
$\Sp(E)\propto k^{\alpha_E}$ with $\alpha_E=\alpha-1$.
As already alluded to above, here, and also in the runs with a Kazantsev
spectrum in the inertial range, $\Sp(B)$ shows a marked decline with $k$.
It is surprising to have such a strong difference in the spectral
properties between $E$ and $B$.
This is probably explained by the mutual cancelation of opposite
parities along the line of sight, which can only affect the parity-odd
$B$ polarization, and would have been missed if one just looked at the
Stokes $Q$ and $U$ polarizations.

\begin{table*}
\centering
\caption{
Diagnostic properties of Runs~A--E.
Except for run C, which has not saturated, the first and second lines
for each run represent the kinematic and saturated phases, respectively.
The characterization of elongated structures concerns the synchrotron
$E$ polarization and is always absent for dust polarization.
On the right, skewness and excess kurtosis are given for $E$ and $B$
during the kinematic and saturated stages.
}\label{Tadd}
\begin{tabular}{cccccllrrrrrrrr} 
\hline
   &     &      &      &      &            &         & \multicolumn{2}{c}{$\SKEW(E)$} & \multicolumn{2}{c}{$\SKEW(B)$} & \multicolumn{2}{c}{$\KURT(E)$} & \multicolumn{2}{c}{$\KURT(B)$} \\
Run& $q_{\rm LS}^{\rm s}$ & $q_{\rm SS}^{\rm s}$ & $q_{\rm LS}^{\rm d}$ & $q_{\rm SS}^{\rm d}$ & elong.\ struct. & spectra & sync & dust & sync & dust & sync & dust & sync & dust \\
\hline
A & 0.84 & 1.41 & 0.80 & 1.04 & no; random & Batchelor             & 0.05 & $-0.01$ &   0.00  & $-0.00$ & 0.20 &   0.00  &   0.17  &   0.00  \\ 
  & 0.87 & 1.18 & 0.80 & 0.76 & no; random & Saffman               & 0.06 & $-0.00$ & $-0.00$ &   0.00  & 0.05 & $-0.00$ & $-0.07$ & $-0.08$ \\ 
B & 0.66 & 3.3  & 0.48 & 1.13 & marginal   & $k^3$ and Saffman     & 0.36 & $-0.03$ &   0.02  &   0.02  & 0.88 &   0.02  &   0.34  & $-0.03$ \\ 
  & 0.68 & 6.4  & 0.53 & 1.34 & marginal   & $k^2$                 & 0.19 & $-0.05$ & $-0.00$ & $-0.01$ & 0.20 &   0.01  &   0.03  & $-0.00$ \\ 
C & 0.42 & 4.6  & 0.39 & 1.05 & weakly     & $k^3$ and short Kaz.  & 1.18 & $-0.09$ &   0.03  & $-0.03$ & 5.94 &   0.04  &   0.41  &   0.07  \\ 
D & 0.37 & 5.7  & 0.34 & 1.14 & very clear & $k^4$ and Kazantsev   & 1.53 & $-0.12$ & $-0.20$ & $-0.01$ & 7.66 &   0.01  &   0.39  &   0.12  \\ 
  & 0.55 & 8.7  & 0.54 & 1.52 & larger scale & $k^2$ and flat part & 0.17 &   0.02  &   0.26  &   0.18  & 0.18 & $-0.07$ &   0.62  &   0.10  \\ 
E & 0.63 & 3.0  & 0.34 & 1.18 & somewhat   & clear Kazantsev       & 0.51 & $-0.03$ &   0.00  &   0.02  & 4.23 &   0.03  &   3.37  &   0.05  \\ 
  & 0.82 & 70   & 0.57 & 1.61 & larger scale & nearly flat         & 0.40 & $-0.19$ &   0.17  &   0.09  & 0.38 & $-0.04$ & $-0.02$ & $-0.19$ \\ 
\hline
\end{tabular}
\end{table*}

It is interesting to note that $\Sp(B)$ shows a strong decline
near $k/k_\nu=0.1$, and it also shows a peak at $\kf$.
This suggests that the $B$ polarization reflects properties of the
velocity field.
The strong decline of $\Sp(B)$ toward large $k$ could therefore be
a signature of the viscous cutoff.

One might have expected that the $E$ and $B$ spectra, which are quadratic
in the magnetic field, have their peak at twice the wavenumber of the
magnetic field spectra.
This expectation was motivated by the fact that magnetic stress
spectra occur at twice the peak wavenumber of the magnetic field itself
\citep{BB20}.
In particular, the peak of $\Sp(\BB^2)$ is twice that of $\Sp(\BB)$.
This is not really seen in the present runs.
Investigating the theoretical relations between the spectra of ${\cal B}$
and ${\cal B}^2$, for example, is clearly of interest, but beyond the
scope of the present paper.

\begin{figure}\begin{center}
\includegraphics[width=\columnwidth]{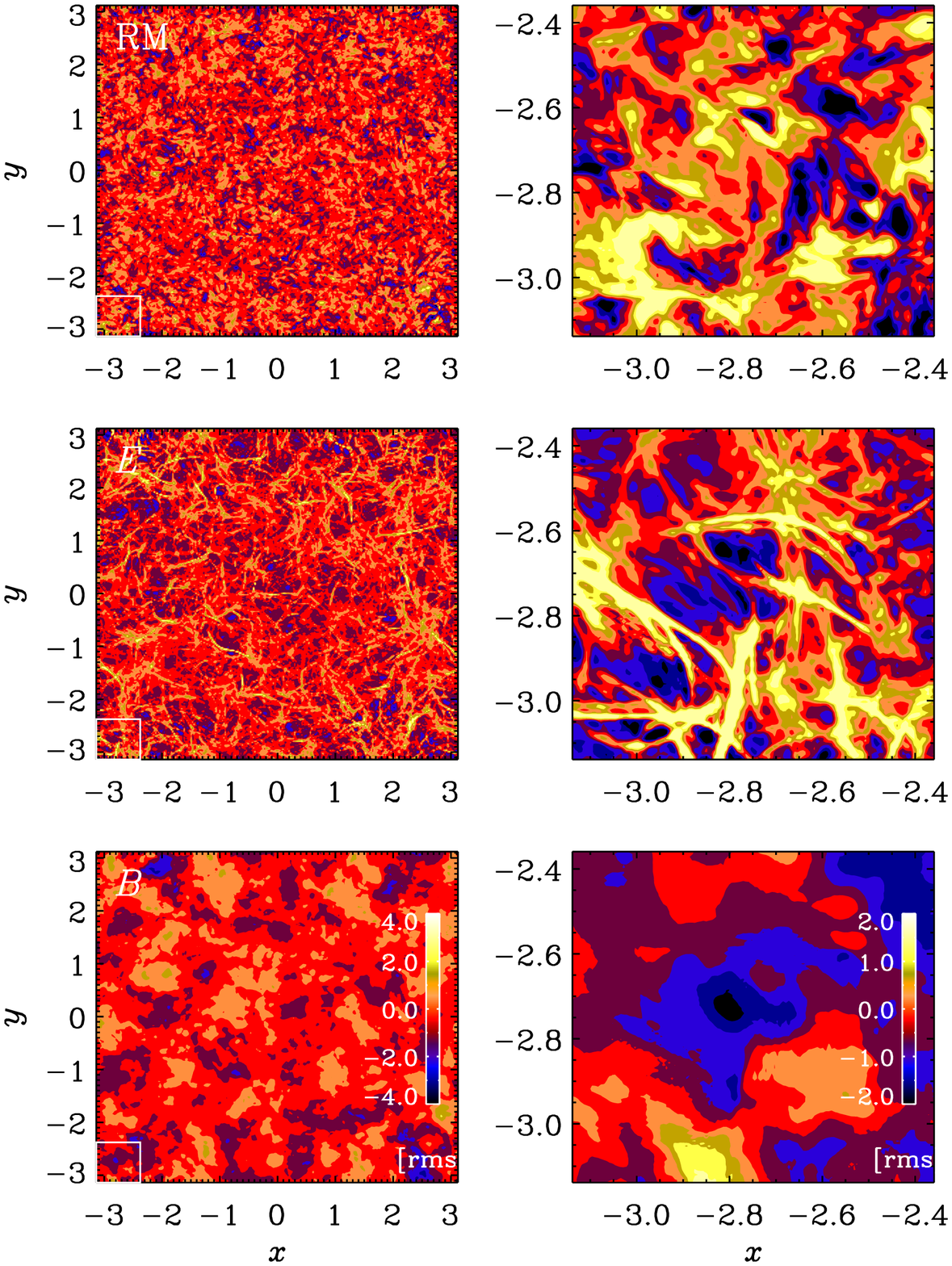}
\end{center}\caption[]{
Same as \Fig{pdiag_planes}, but for Run~D during the kinematic stage.
}\label{pdiag_planes_D1024_Pm30a_kf4}\end{figure}

\begin{figure}\begin{center}
\includegraphics[width=\columnwidth]{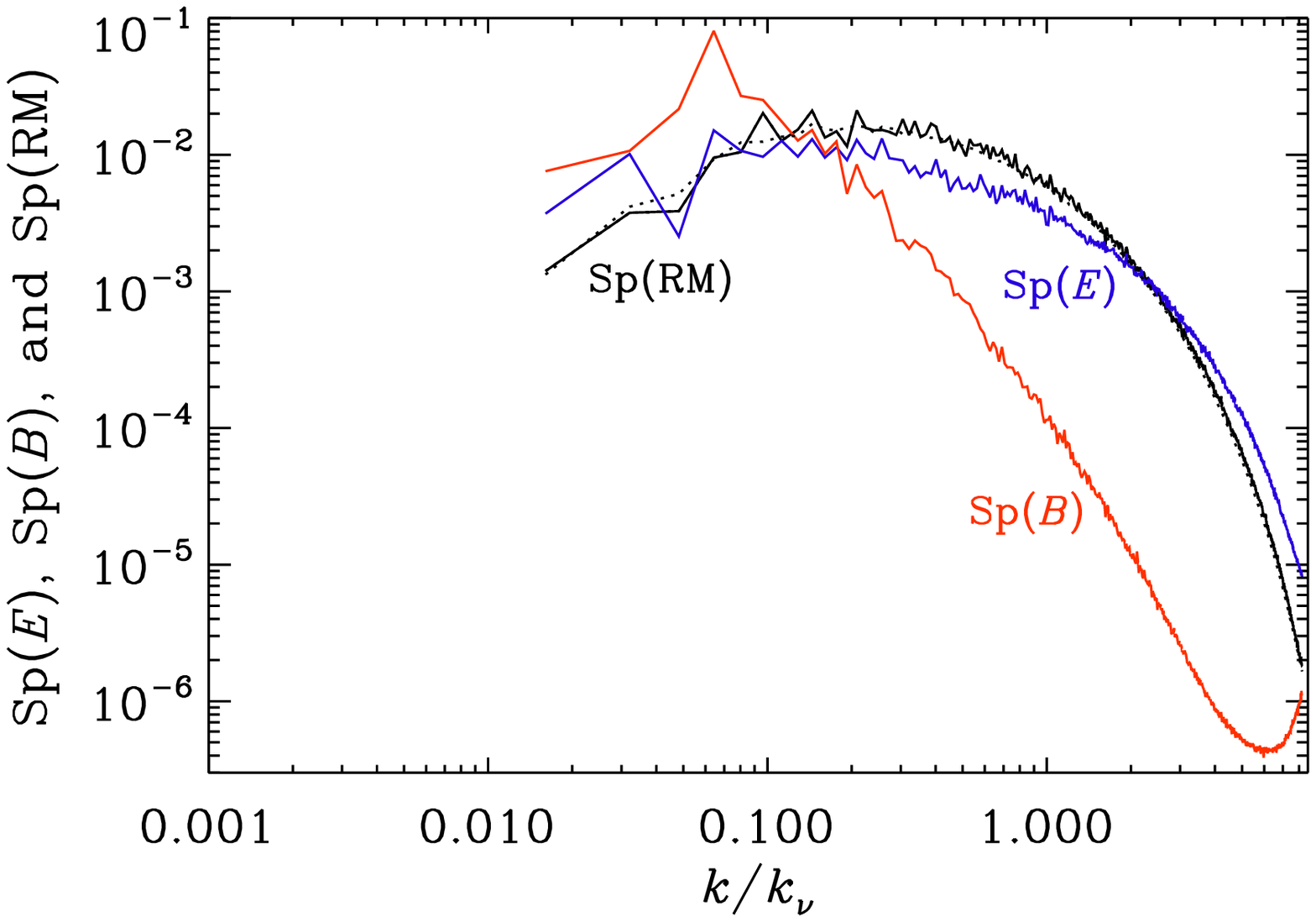}
\end{center}\caption[]{
Same as \Fig{pdiag_spec}, but for Run~D during the kinematic stage.
}\label{pdiag_spec_D1024_Pm30a_kf4}\end{figure}

For Run~D, which has a large magnetic Prandtl number, we see very
pronounced elongated structures in the $E$ polarization, which is
not seen in the $B$ polarization, and only to some extent in RM;
see \Fig{pdiag_planes_D1024_Pm30a_kf4}.
This corresponds with the spectra shown in
\Fig{pdiag_spec_D1024_Pm30a_kf4}, where $\Sp({\rm RM})$ and $\Sp(E)$ have
a similar shape, but $\Sp(B)$ shows a sharp decline with increasing $k$.

\subsection{Comparison with dust polarization}

\begin{figure}\begin{center}
\includegraphics[width=\columnwidth]{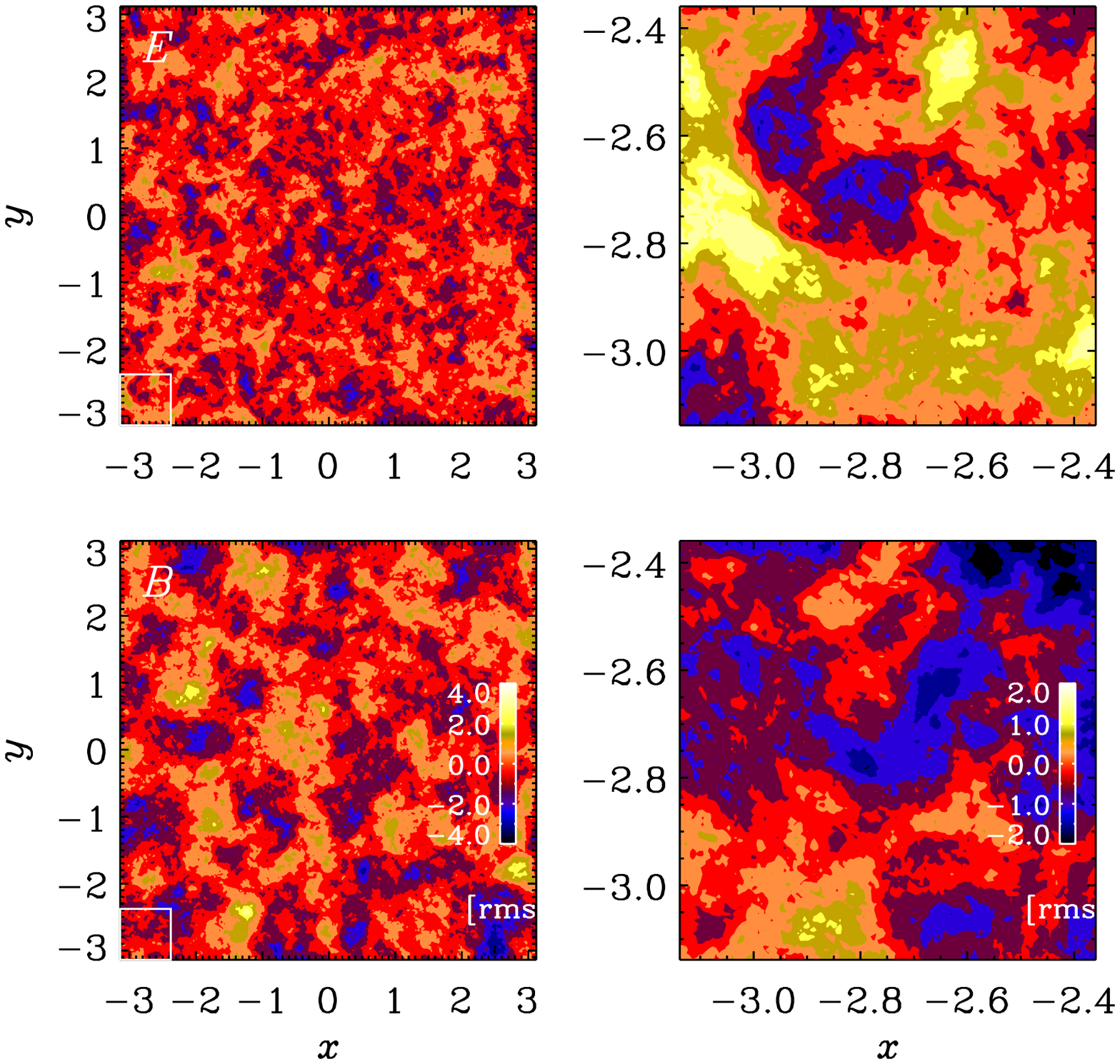}
\end{center}\caption[]{
$E$ and $B$ for dust polarization from Run~D.
}\label{pdiag_planes_DUST_D1024_Pm30a_kf4}\end{figure}

We recall that the main difference between synchrotron and dust emission
lies in the fact that dust emission depends mostly on the dust temperature
and not on the magnetic field strength \citep{PlanckXX, Bracco+19}.
For dust emission, therefore, weak fields contribute just as much as
strong fields.
This is in sharp contrast to synchrotron emission, where the emissivity
scales approximately quadratically with the magnetic field strength
\citep{Ginzburg+Syrovatskii65}.
This causes systematic differences between the spectra from dust and
synchrotron emission.
Most remarkably, the elongated structures seen in
the $E$ polarization of synchrotron emission are now
absent; compare \Fig{pdiag_planes_D1024_Pm30a_kf4} with
\Fig{pdiag_planes_DUST_D1024_Pm30a_kf4}.
Comparing the spectra in \Figs{pdiag_spec_D1024_Pm30a_kf4}
{pdiag_spec_DUST_D1024_Pm30a_kf4}, we see that the excess
power in $E$ is now absent.
This is quantified in more detail in the next section.

\begin{figure}\begin{center}
\includegraphics[width=\columnwidth]{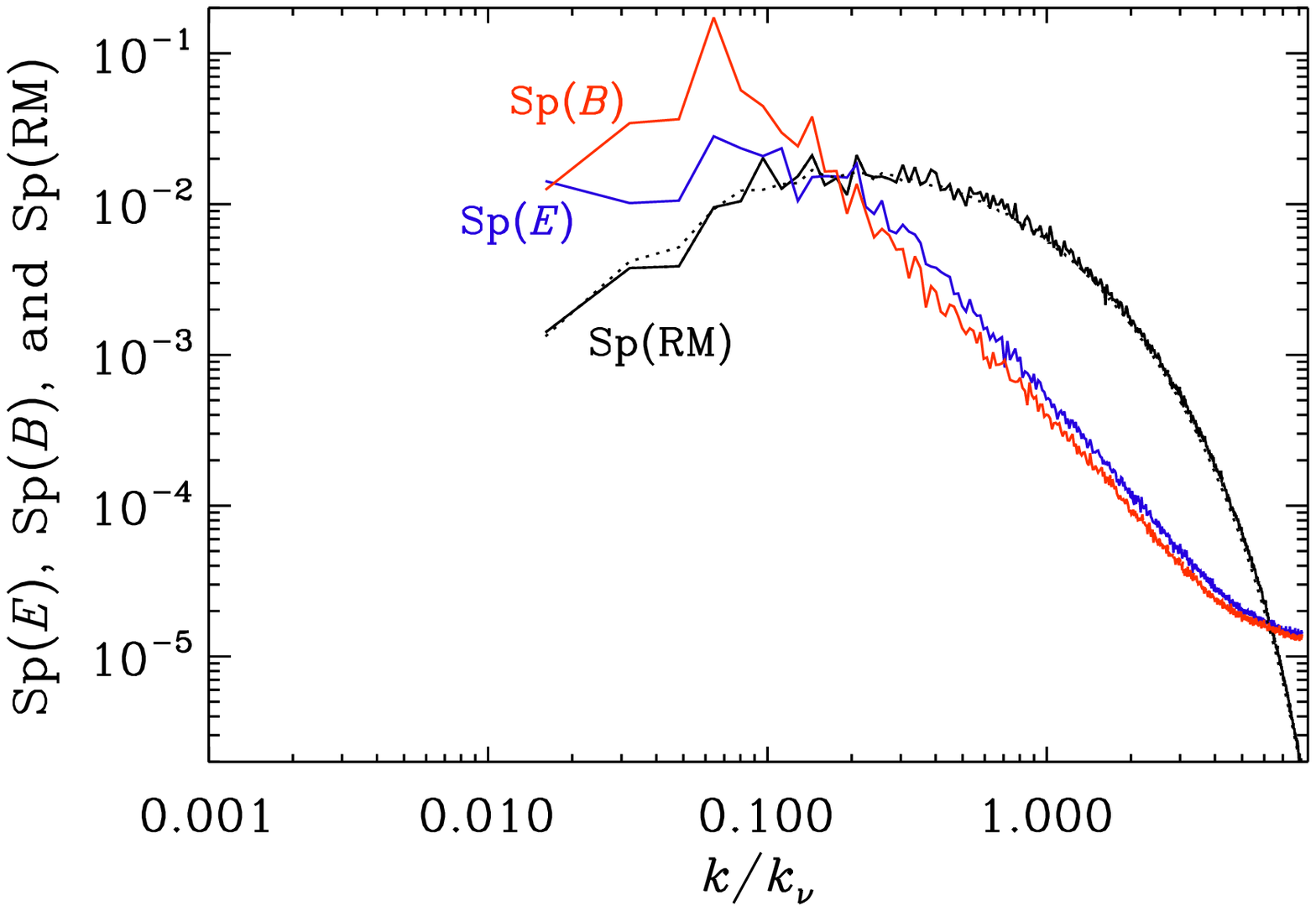}
\end{center}\caption[]{
Same as \Fig{pdiag_spec_D1024_Pm30a_kf4}, but for dust polarization.
}\label{pdiag_spec_DUST_D1024_Pm30a_kf4}\end{figure}

\subsection{Excess $E$ polarization in statistics}

Of particular interest is the ratio $\bra{E^2}/\bra{B^2}$.
As we saw from \Fig{pdiag_spec}, the answer may depend on the
wavenumber range for which the data is taken.
For Runs~D and E, the $E$ and $B$ spectra cross,
and the crossing point $k_\times$ of the $E$ and $B$ spectra
lies at $k_\times/k_\nu\approx0.14$.
It is therefore useful to compute the ratio separately for small and
large $k$.
\begin{equation}
q_{\rm LS}\equiv\bra{E^2}_{\rm LS}/\bra{B^2}_{\rm LS}=
\left.\int_0^{k_\times}\Sp(E)\,\dd k\right/\int_0^{k_\times}\Sp(B)\,\dd k,
\end{equation}
\begin{equation}
q_{\rm SS}\equiv\bra{E^2}_{\rm SS}/\bra{B^2}_{\rm SS}=
\left.\int_{k_\times}^\infty\Sp(E)\,\dd k\right/\int_{k_\times}^\infty\Sp(B)\,\dd k,
\end{equation}
To distinguish the ratios for synchrotron and dust emission,
we add superscripts s and d, respectively.
The resulting ratios are listed in \Tab{Tadd} both for synchrotron
emission ($q_{\rm LS}^{\rm s}$ and $q_{\rm SS}^{\rm s}$) and for dust
emission ($q_{\rm LS}^{\rm d}$ and $q_{\rm SS}^{\rm d}$), along with
other observations about the runs.
For synchrotron emission, the values of $q_{\rm SS}^{\rm s}$ can be
rather large compared with the aforementioned factor of two.
For dust polarization, the values are significantly smaller, although
values of $q_{\rm SS}^{\rm d}$ of 1.5 and 1.6 can be seen for Runs~D
and E, respectively.

\begin{figure}\begin{center}
\includegraphics[width=\columnwidth]{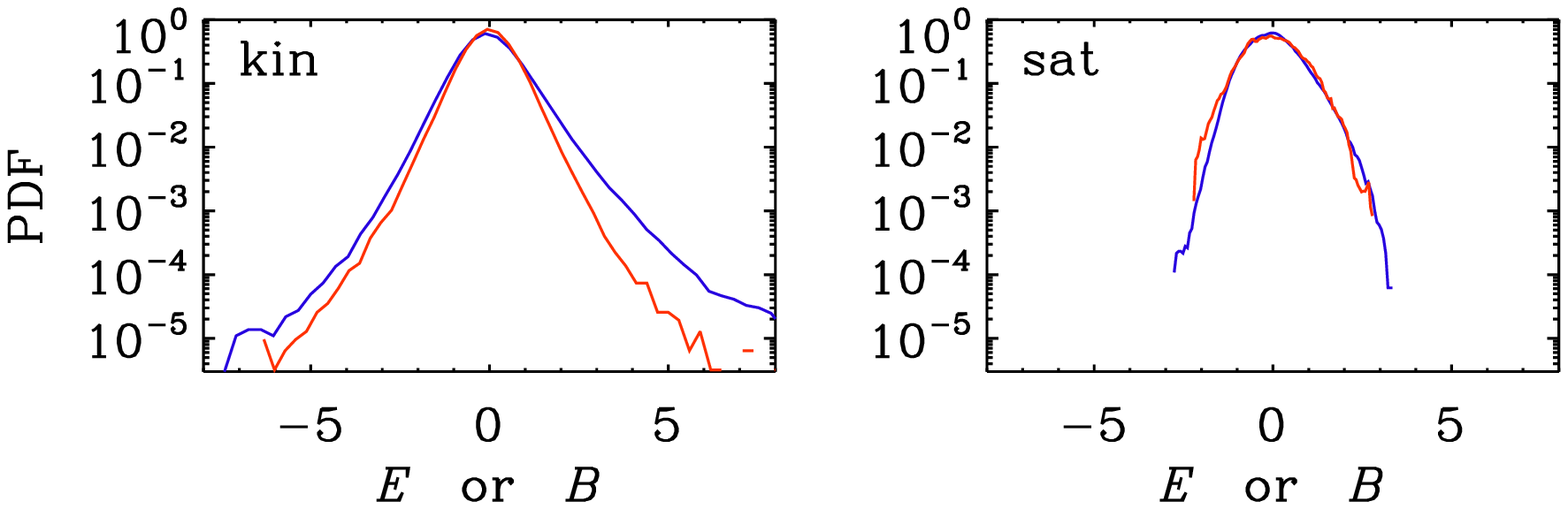}
\end{center}\caption[]{
PDFs of $E$ (blue lines) and $B$ (red lines) during the kinematic (left)
and saturated (right) stages for Run~E.
}\label{pdiag_pdf_D1024_Pm1c_kf4}\end{figure}

\begin{figure}\begin{center}
\includegraphics[width=\columnwidth]{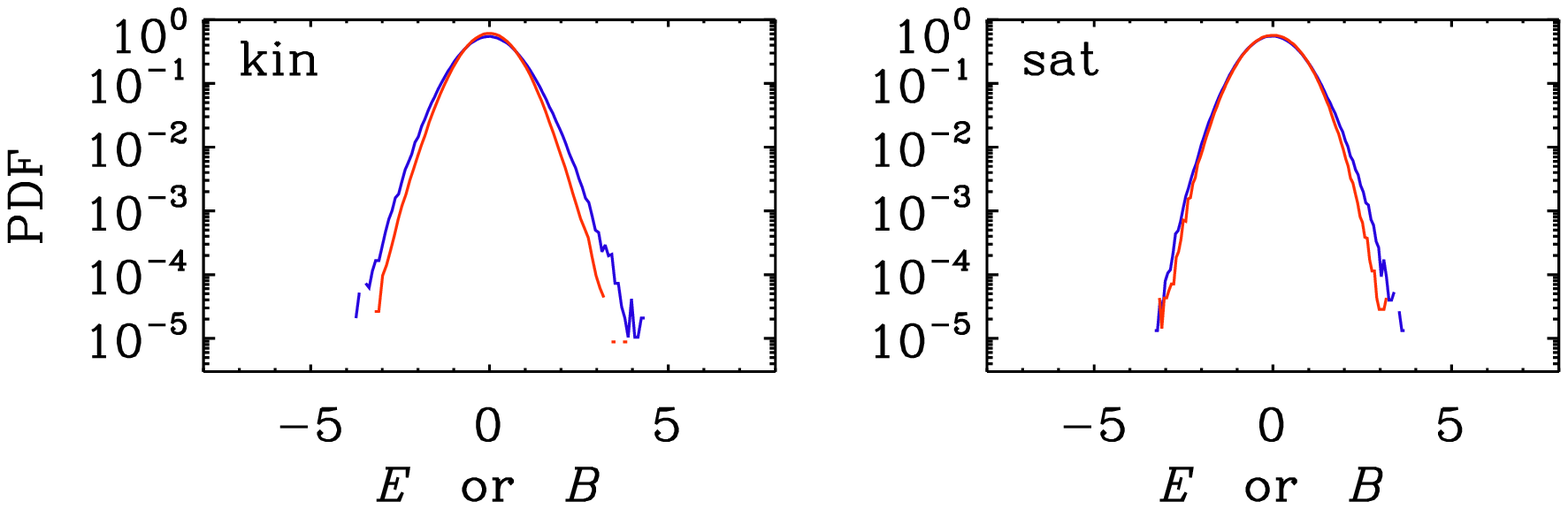}
\end{center}\caption[]{
Same as \Fig{pdiag_pdf_D1024_Pm1c_kf4}, but for Run~A.
}\label{pdiag_pdf_D1024_Pm1e_kf120}\end{figure}

\begin{figure}\begin{center}
\includegraphics[width=\columnwidth]{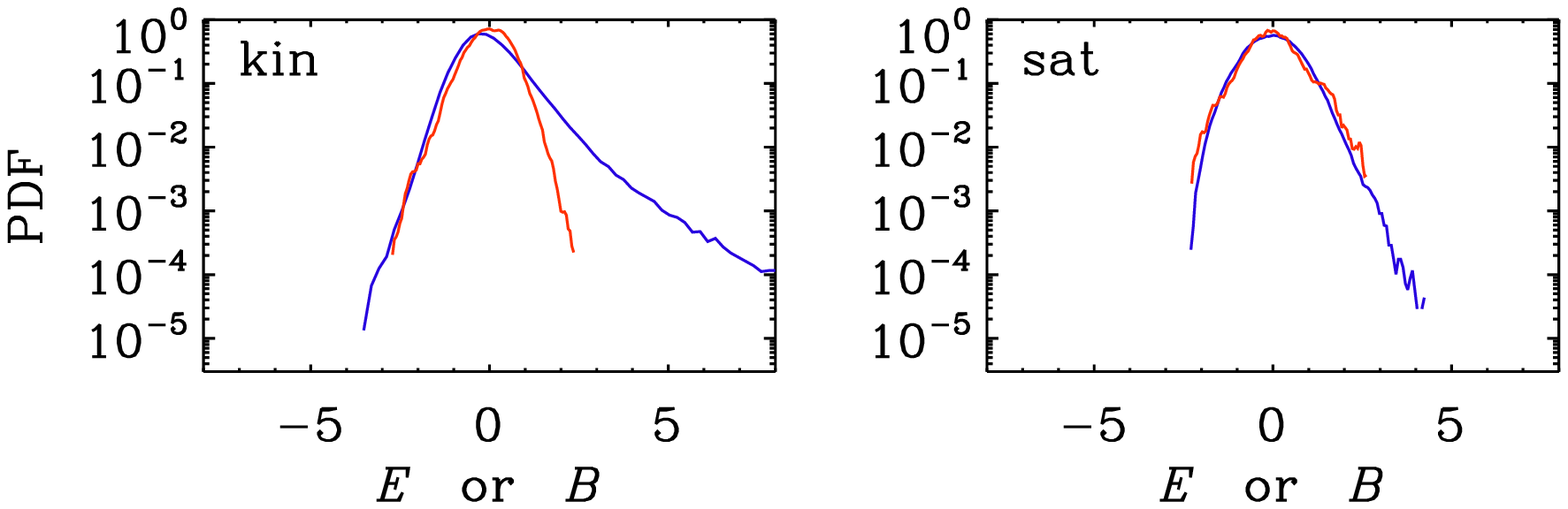}
\end{center}\caption[]{
Same as \Fig{pdiag_pdf_D1024_Pm1c_kf4}, but for Run~D.
}\label{pdiag_pdf_D1024_Pm30a_kf4}\end{figure}

Earlier work on the $E$ and $B$ polarizations has shown a tendency for
the probability functions (PDFs) of $E$ to be non-Gaussian and skewed,
while the $B$ polarization was more nearly Gaussian \citep{Bra+19b,Bra19},
especially in decaying turbulence.
In the present case, the result depends on the existence of an inertial
range (Run~E) and on whether the run is saturated or not.
In \Fig{pdiag_pdf_D1024_Pm1c_kf4} we show that for Run~E, the PDFs
correspond to stretched exponentials during the kinematic stage,
with $E$ being also skewed, but both become nearly Gaussian during the
saturated stage.
For Run~A, on the other hand, $E$ and $B$ are nearly Gaussian
both during the kinematic and the saturated stages; see
\Fig{pdiag_pdf_D1024_Pm1e_kf120}.
Run~D is closer to Run~E than to Run~A, but with the $E$
polarization being even more skewed during the kinematic stage;
see \Fig{pdiag_pdf_D1024_Pm30a_kf4}.
This could be a signature of the large magnetic Prandtl number in
this case.
In the second part of \Tab{Tadd}, we summarize the resulting values for skewness and
(excess) kurtosis for all runs during the kinematic and saturated stages.
For synchrotron emission,
both the kurtosis and the skewness are particularly high for the runs with
large magnetic Prandtl number (Runs~C and D) during the kinematic phase.
For Run~E, the kurtosis of $E$ is also fairly large during the
kinematic phase, but the skewness is now only 0.5.
For this run, furthermore, even the kurtosis of $B$ is significant.
For dust emission, on the other hand, both skewness and excess kurtosis
are relatively small.

For completeness, we present images and spectra of $E$, $B$, and RM for
the other runs in \App{Others} and in the supplemental material on Zenodo.
The magnetic field tends to develop larger scale structures in the
saturated state, which is also shown in this appendix.

\section{Non-isothermal two-phase flows}
\label{TwoPhase}

To assess the robustness of our results to the assumption of an
isothermal equation of state, we now consider a simulation
with an ideal equation of state instead.
This means that we must also include an evolution equation for
the specific entropy, $s$, namely,
\begin{equation}
T{\DD s\over\DD t}=2\nu\SSSS^2
+{1\over\rho}\nab\cdot\left(\cp\rho\chi\nab T\right)-{\cal L},
\end{equation}
where $T$ is the temperature, $\chi$ is the thermal diffusivity,
$\cp$ is the specific heat at constant pressure,
and ${\cal L}$ is the net cooling,
\begin{equation}
{\cal L}=\rho\Lambda-\Gamma,
\end{equation}
where $\Gamma=\mbox{const}$ is assumed for the heating function
and $\Lambda(T)$ is the cooling function.

For the following model, we take the forced turbulence
simulation setup of \cite{BKM07}; see their Section~3.3.
They adopted the piecewise power-law parameterization of \cite{SS02}
for $\Lambda(T)$, but with slightly modified coefficients so as to
avoid discontinuities; see Table~1 of \cite{BKM07}.
Also, the $\cs^2\nab\ln\rho$ term in \Eq{DuDt} now includes
the specific entropy gradient, i.e., we replace
\begin{equation}
\cs^2\nab\ln\rho\quad\longrightarrow\quad \cs^2\nab(\ln\rho+s/\cp),
\end{equation}
where $\cs^2=(\gamma-1)\,\cp T$ is now no longer constant.
Here, $\gamma=\cp/\cv=5/3$ is the ratio of specific heats,
with $\cv$ being the specific heat at constant volume.
Note that $\cs$ is related to $\ln\rho$ and $s$ via
$\cs^2=\csz^2[\gamma s/\cp+(\gamma-1)\ln(\rho/\rho_0)]$,
where $\csz$ and $\rho_0$ are constants.
\cite{BKM07} chose $\csz=1\kms$ and $\rho_0=1\,m_{\rm H}\cm^{-3}$,
where $m_{\rm H}$ is the mass of the hydrogen atom.
They also chose $k_1=1\kpc^{-1}$, which fixes then the unit of time,
although most of our results are presented in nondimensional form.
The adopted cooling function allows for two stable fixed points
of $\cs\approx1\kms$ and $\approx10\kms$.
When the initial density is in a suitable range (here $\rho=\rho_0$
initially) in the flow segregates into two phases, which is why we
talk about a two-phase flow.
Dynamos in such flows have recently been studied by \cite{SF22},
who found a slight suppression of dynamo action due to the
presence of two phases.
We refer to our two-phase models as Run~T0 and T1, where the
numeral indicates the absence or presence of helicity in the
forcing function, although there are several other difference
between those two runs as well.
Since we only consider the kinematic phase of the dynamo,
the presence of helicity does not play an important role,
because the large-scale dynamo would only emerge during
saturation \citep{Bra01}.

\begin{figure}\begin{center}
\includegraphics[width=\columnwidth]{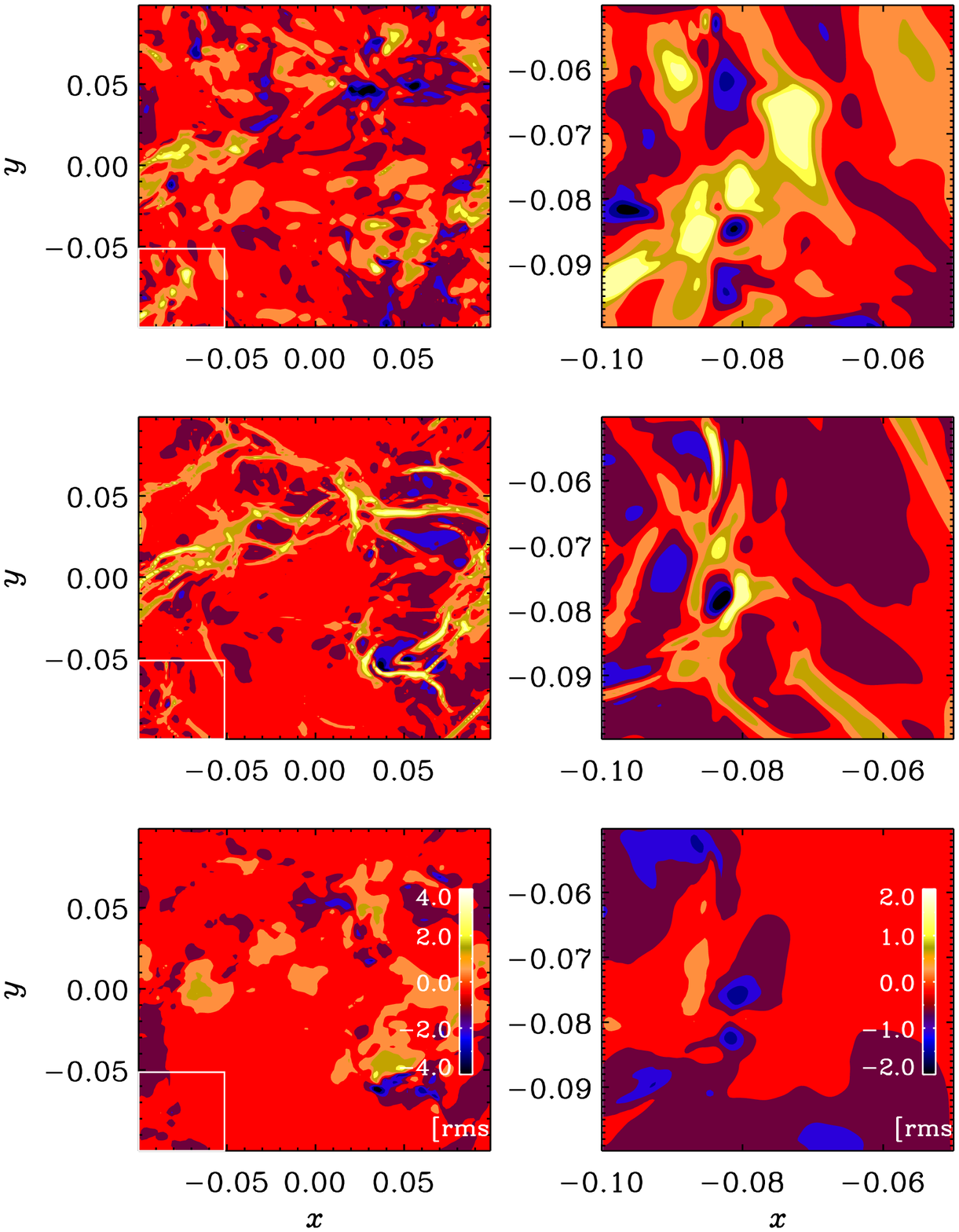}
\end{center}\caption[]{
Same as \Fig{pdiag_planes}, but for Run~T0 during the kinematic stage.
}\label{pdiag_planes_Mfor512c4}\end{figure}

\begin{figure}\begin{center}
\includegraphics[width=\columnwidth]{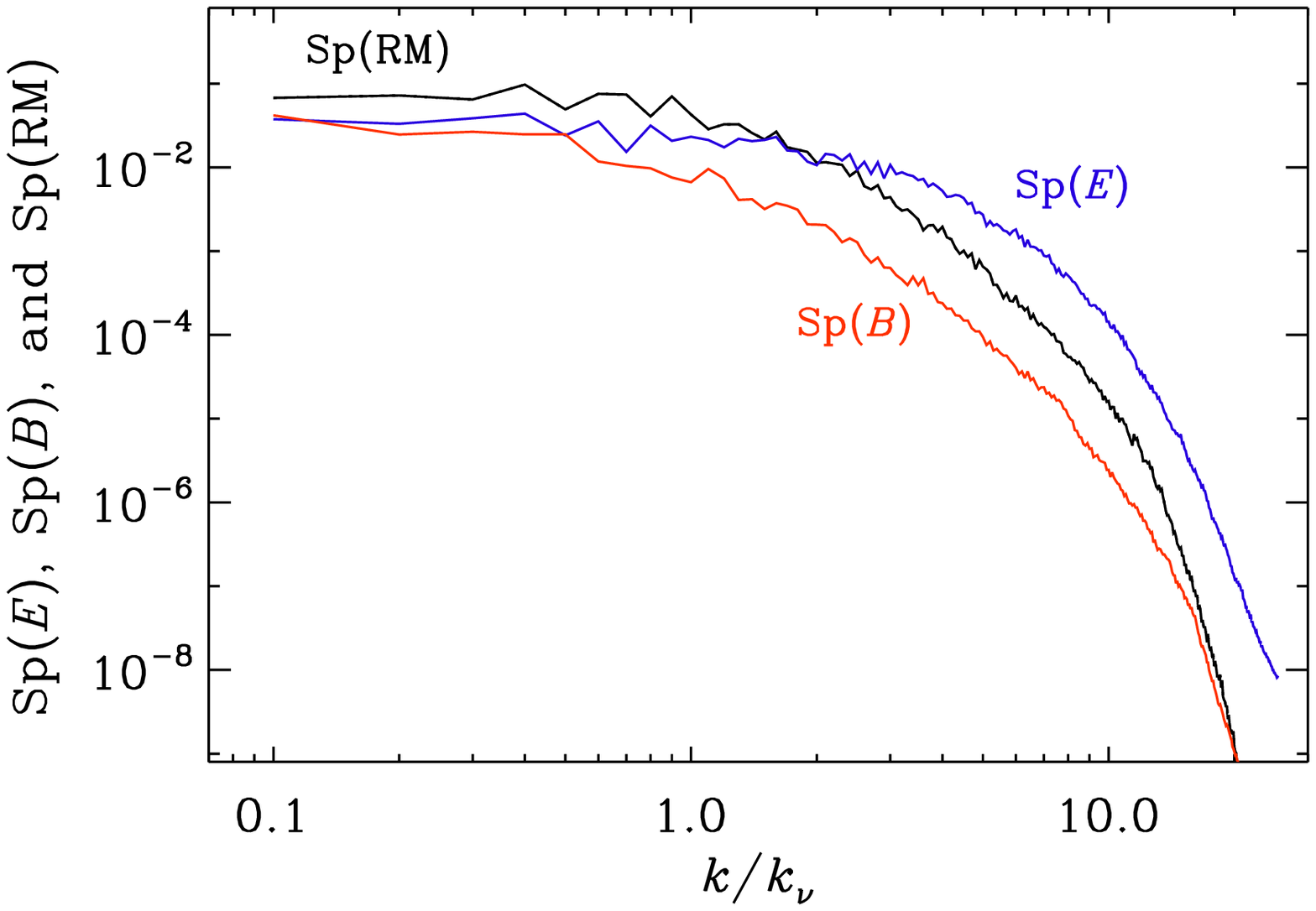}
\end{center}\caption[]{
Same as \Fig{pdiag_spec}, but for Run~T0 during the kinematic stage.
}\label{pdiag_spec_Mfor512c4}\end{figure}

The variability of the sound speed implies that in cool regions, the
flow can become highly supersonic.
On the average, however, the Mach number is around unity.
The possibility of large Mach numbers requires the use of large
viscosity and large thermal and magnetic diffusivities.
Also, the simulations of \cite{BKM07} employed a helical forcing
function, which helps lowering the threshold for dynamo action.
We denote the fractional helicity of the forcing by $\sigma$,
so $\sigma=1$ for Run~T1, but we also present Run~T0 with $\sigma=0$.
Thus, our non-isothermal simulations are in many ways quite different
from those presented in the rest of the paper and deserve proper analysis
in a separate paper.
Nevertheless, it is important to point out that the resulting images for
RM, $E$ and $B$, shown in \Fig{pdiag_planes_Mfor512c4} for Run~T0,
are similar to those shown earlier in the paper.
Also the spectra shown in \Fig{pdiag_spec_Mfor512c4} are similar,
except that a sharp peak at the forcing scale is absent in $\Sp(B)$.
The corresponding plots for Run~T1 are shown in the supplemental material.

\begin{table}
\centering
\caption{
Summary of the parameters for Runs~T0 and T1.
}\label{TRunT}
\begin{tabular}{ccccccccc} 
\hline
    &          & Ma       & & & & & & \\
Run & $\sigma$ & rms, max & $\tilde{k}_{\rm f}$ & $\tilde{k}_\nu$ &
$\tilde{\gamma}$ & $\Rey$ & $\Rm$ & $\Pm$ \\
\hline
T1 & 1 & 8, 40 & 3 & 11 & 0.003 & 130 & 3800 & 30 \\ 
T0 & 0 & 5, 20 & 4 & 10 & 0.001 & 150 & 4600 & 30 \\ 
\hline
\end{tabular}
\end{table}

In this model, $\Pm=10$; see \Tab{TRunT} for a summary of parameters for
this model.
Consistent with the fact that the magnetic Prandtl number here is larger
than unity, we find also here that the PDFs are skewed similarly as for
Run~D in \Fig{pdiag_pdf_D1024_Pm30a_kf4}.
In addition to the different values of $\sigma$ for Runs~T0 and T1,
we used a five times smaller forcing amplitude for Run~T0, but
the Mach numbers are not so different.
This is presumably caused by the two-phase nature of the flow.
Indeed, it has been argued that two-phase flows can lead to sustained
turbulence \citep{Iwasaki+Inutsuka14,Kobayashi+20}, but clarifying this
conclusively may require larger resolution and comparison with cases
where the dynamical viscosity, $\rho\nu$, is held constant.

\section{Conclusions}

It is well known that in virtually all cases of astrophysical interest,
Kolmogorov-type turbulence is always accompanied by dynamo action.
This has consequences for the way turbulent energy is being dissipated
into heat and radiation, which depends strongly on the value of the
magnetic Prandtl number \citep{Bra14,BR19}.
It is also well known that in the kinematic regime, the small-scale
dynamo produces a characteristic spectrum known as the Kazantsev
spectrum \citep{Kaz68}, which was later discussed in more detail by
\cite{Kulsrud+Anderson92}.
The Kazantsev spectrum is now also clearly seen in simulations
\citep{Scheko04, HBD04}.
Our work has now shown that its spectrum extends over the full inertial
range of the turbulence and that on larger, subinertial scales, one has
a Batchelor spectrum, which turns into a Saffman spectrum as the dynamo
saturates.

The fact that the Kazantsev spectrum extends over the full inertial
range and not just over the subviscous range, $k_\nu\ll k\ll k_\eta$,
is worth highlighting.
Owing to limited numerical resolution, $k_\nu$ was in earlier numerical
simulations often too close to the forcing wavenumber $\kf$.
This reinforced the theoretical expectation that small-scale dynamo
action is confined to the wavenumber range where the flow is smooth
\citep{Scheko04}.
On the other hand, in the inertial range, where the flow is ``rough,''
small-scale dynamos should still work, but they are much harder to
excite \citep{RK97}.
This is the case at small $\Pm$, when $k_\eta<k_\nu$ and the magnetic
energy spectrum peaks in the inertial range.
Since the work of \cite{Iskakov+07} we know that small-scale dynamos
do indeed work for $\Pm\ll1$.
A practical difficulty here lies in the fact that between the viscous
and inertial subranges, there is the bottleneck range \citep{Falk94},
where turbulence is even rougher than in the inertial range \citep{BC04},
making dynamo action even harder to demonstrate.
This is not a problem in the magnetically saturated case, because
then the bottleneck is suppressed \citep{Bra11}.
In any case, for our Run~E, we have $\Rm\approx1600$, which should be
large enough for the small-scale dynamo to be excited over the whole
inertial range.
Therefore, one should not be too surprised if the Kazantsev spectrum
does indeed extend throughout the entire inertial range of the turbulence.

For synchrotron emission, our results suggest an excess $E$
polarization over the $B$ polarization at subresistive scales,
and the opposite trend is found at larger scales.
This is also found for the non-isothermal two-phase flows discussed in
\Sec{TwoPhase}.
For dust emission, on the other hand, the subviscous $E$ excess is much
weaker, although values of 1.5 and 1.6 can be found for large magnetic
Reynolds numbers.

An excess of the $E$ polarization is observed in the Galactic microwave
foreground emission \citep{PlanckXXX, Caldwell}.
One might have expected the relevant
scales to be much larger than the viscous scales.
Observationally, however, one cannot exclude the possibility that
the observed excess could result from subviscous scale.
Furthermore, even in the saturated case, when the Kazantsev spectrum
has disappeared, there is still an excess at subviscous scales.
This numerical finding seems therefore remarkably robust.

Our study motivates new targets of investigation and new questions.
How generic are the different realizations of turbulence found in the
present study?
Can we really expect the modeled types of velocity and magnetic fields
to occur in galaxy clusters or in the ISM?
One reason for concern is the fact that in all our flows, the driving
is monochromatic with a typical wavenumber $\kf$.
Real turbulence may be more complicated.
Nevertheless, the turbulence should always be characterized by a typical
energy-carrying scale, $\xiM$, which defines an approximate position
of the spectral peak at $\kf\approx\xiM^{-1}$.
It is therefore not obvious, that the monochromatic driving of our
turbulence is actually very restrictive.

It is remarkable that the existence of the Kazantsev spectrum appears
to be fairly insensitive to the value of the magnetic Prandtl number,
but it never occurs at wavenumbers below the turbulent inertial range.
Observing the transition to the steeper Batchelor spectrum requires
very large domain sizes.
This is why we allowed for a forcing scale that was up to 120 times
shorter than the size of the domain (Run~A).
On the other hand, it is conceivable that in real cluster turbulence,
the subinertial range is not entirely free of driving, as was assumed
in the present work.
However, clarifying this observationally could be difficult and may
require full-sky observations to be able to identify the true peak
of the spectrum.

Real galaxy cluster turbulence is believed to be driven by cluster mergers
\citep[e.g.,][]{Roettiger+99}, and that the turbulence would be a state of decay
in between such mergers events.
An open question is therefore whether the Kazantsev spectrum can also be
seen in decaying turbulence.
There is no reason why not, but it is not easy to find and requires,
as we have now seen, a sufficiently extended inertial range.
On the other hand, a large magnetic Prandtl number is not required.

Our work has shown that differences between the parity-even $E$ and the
parity-odd $B$ polarizations may serve to distinguish between Kazantsev
and Batchelor spectra in synchrotron emission, provided the magnetic
field is still (or again) in a kinematic growth phase.
The best chance to find turbulence in a kinematic state may be in clusters
in the beginning of a merger event, as alluded to above.
However, our present work would need to be adapted to such situations
to develop more realistic observational signatures specific to clusters
turbulence.

All our runs with noticeable spectral differences between $E$ and $B$
had Kazantsev spectra in the inertial range.
There were also cases where the PDFs were non-Gaussian with strong
skewness in the $E$ polarization.
Again, this does not occur for dust emission.
Subviscous scales played a decisive role in producing excess $E$
polarization.
It is unclear whether those are observationally accessible.
Of course, if the observed excess $E$ polarization can only
be explained as a subviscous phenomenon, it might just be this effect
that would give us information about subviscous scales.
Simulations by \cite{Kri+18} for dust emission also produced excess $E$
polarization, but the reason behind this was not clear.
Those simulations where ideal ones, so the viscous scale was an entirely
numerical phenomenon in their simulations.
Nevertheless, a more detailed spectral analysis might help shedding more
light on the phenomenon of excess $E$ polarization.
The strong skewness in the synchrotron $E$ polarization during the
kinematic regime may have been caused by the elongated structures that
are also clearly seen in images of $E$, provided $\Pm$ is large.
This is expected to be the case in the ISM and in galaxy clusters,
which motivates further morphological and statistical studies of
observed $E$ and $B$ polarizations.
It is also striking that the $B$ polarization tends to reflect
the velocity field and shows a peak at the driving scale.
The $E$ polarization, by contrast, tends to peak at the wavenumber
where the magnetic field is strong.
This difference clearly sticks out in images showing larger scale
structures of $B$ than those of $E$.
This could be another characteristic signature detectable in the ISM.

\section*{Acknowledgements}

We thank Robi Banerjee for the suggestion to discuss the effect of a
non-isothermal equation of state.
This work emerged during discussions at the Nordita program
on ``Magnetic field evolution in low density or strongly
stratified plasmas'' in May 2022.
The research was supported by the Swedish Research Council
(Vetenskapsr{\aa}det, 2019-04234).
Nordita is sponsored by Nordforsk.
We acknowledge the allocation of computing resources provided by the
Swedish National Allocations Committee at the Center for Parallel
Computers at the Royal Institute of Technology in Stockholm and
Link\"oping.

\section*{Data Availability}
 
The source code used for the simulations of this study,
the {\sc Pencil Code} \citep{JOSS}, is freely available on
\url{https://github.com/pencil-code/}.
The DOI of the code is https://doi.org/10.5281/zenodo.2315093.
The simulation setups and the corresponding secondary data, as well
as supplemental material with additional plots for the
PDFs of Run~B and diagnostic images and spectra for Runs~B and D
in the saturated state are available on
\url{https://doi.org/10.5281/zenodo.6862459}; see also
\url{https://www.nordita.org/~brandenb/projects/Kazantsev-Subinertial}
for easier access to the same material as on the Zenodo site.



\bibliographystyle{mnras}
\bibliography{ref} 


\newpage
\appendix

\section{Changes during saturation}
\label{Changes}

\begin{table}
\centering
\caption{
Values of $\Ma$, $\tilde{k}_\nu$, and $\tilde{\epsilon}_{\rm K}=\epsK/k_1\cs^3$
during the kinematic and saturated stages.
}\label{Tsat}
\begin{tabular}{cccrrcc} 
\hline
Run & $\Ma$ & sat. & $\tilde{k}_\nu$ & sat. &
$\tilde{\epsilon}_{\rm K}$ & saturated \\
\hline
A & 0.111 & 0.102 & 764 & 718 & $9.2\times10^{-3}$ & $7.2\times10^{-3}$ \\ 
B & 0.121 & 0.105 & 389 & 328 & $2.9\times10^{-3}$ & $1.5\times10^{-3}$ \\ 
C & 0.118 &  ...  & 106 & ... & $9.9\times10^{-4}$ &  ... \\ 
D & 0.122 & 0.079 &  62 &  50 & $4.1\times10^{-4}$ & $1.7\times10^{-4}$ \\ 
E & 0.130 & 0.096 & 461 & 330 & $3.6\times10^{-4}$ & $9.5\times10^{-5}$ \\ 
\hline
\end{tabular}
\end{table}

As the dynamo saturates, $\urms$, $k_\nu$, and $\epsK$ decrease by
a certain amount that depends on the input parameters.
This is demonstrated in \Tab{Tsat}, where we list the kinematic and
saturated values for all five runs.
We recall that Run~C was not continued into saturation, which is
here indicated by the ellipses.

Based on the values listed in \Tab{Tsat}, we can infer that the ratios of
the saturated to the kinematic values depend either on $\Rm$ or on $\Rey$.
Specifically, we see that
$\urms(\sat)/\urms(\kin)$ decreases mainly with $\Rm$ like $\Rm^{-0.07}$,
$k_\nu(\sat)/k_\nu(\kin)$ decreases like $\Rm^{-0.04}$, and
$\epsK(\sat)/\epsK(\kin)$ decreases mainly with $\Rey$ like $\Rey^{-1/4}$.

\section{Diagnostics for other runs}
\label{Others}

\begin{figure}\begin{center}
\includegraphics[width=\columnwidth]{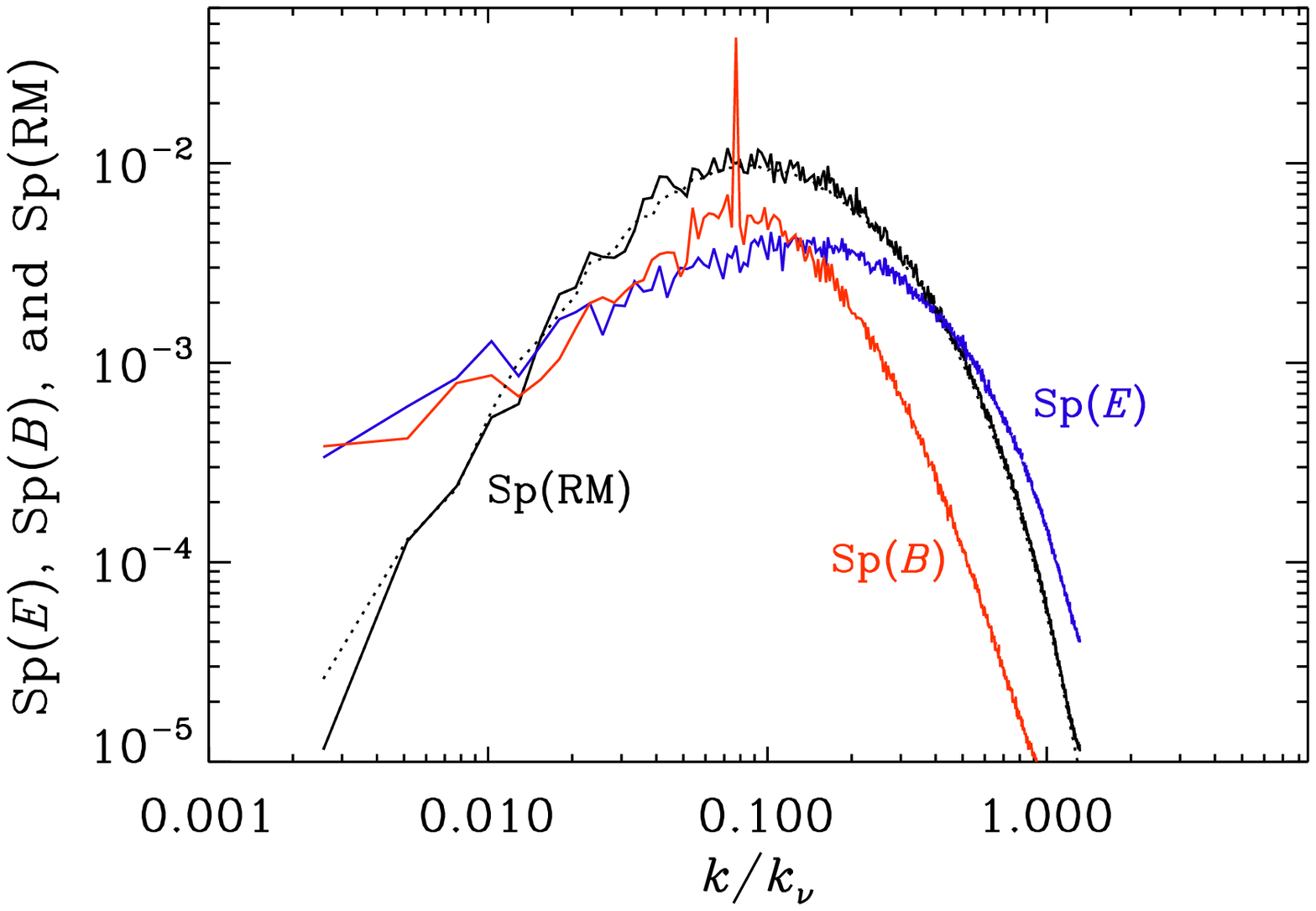}
\end{center}\caption[]{
Same as \Fig{pdiag_spec}, but for Run~B during the kinematic stage.
}\label{pdiag_spec_D1024_Pm1d_kf30}\end{figure}

\begin{figure}\begin{center}
\includegraphics[width=\columnwidth]{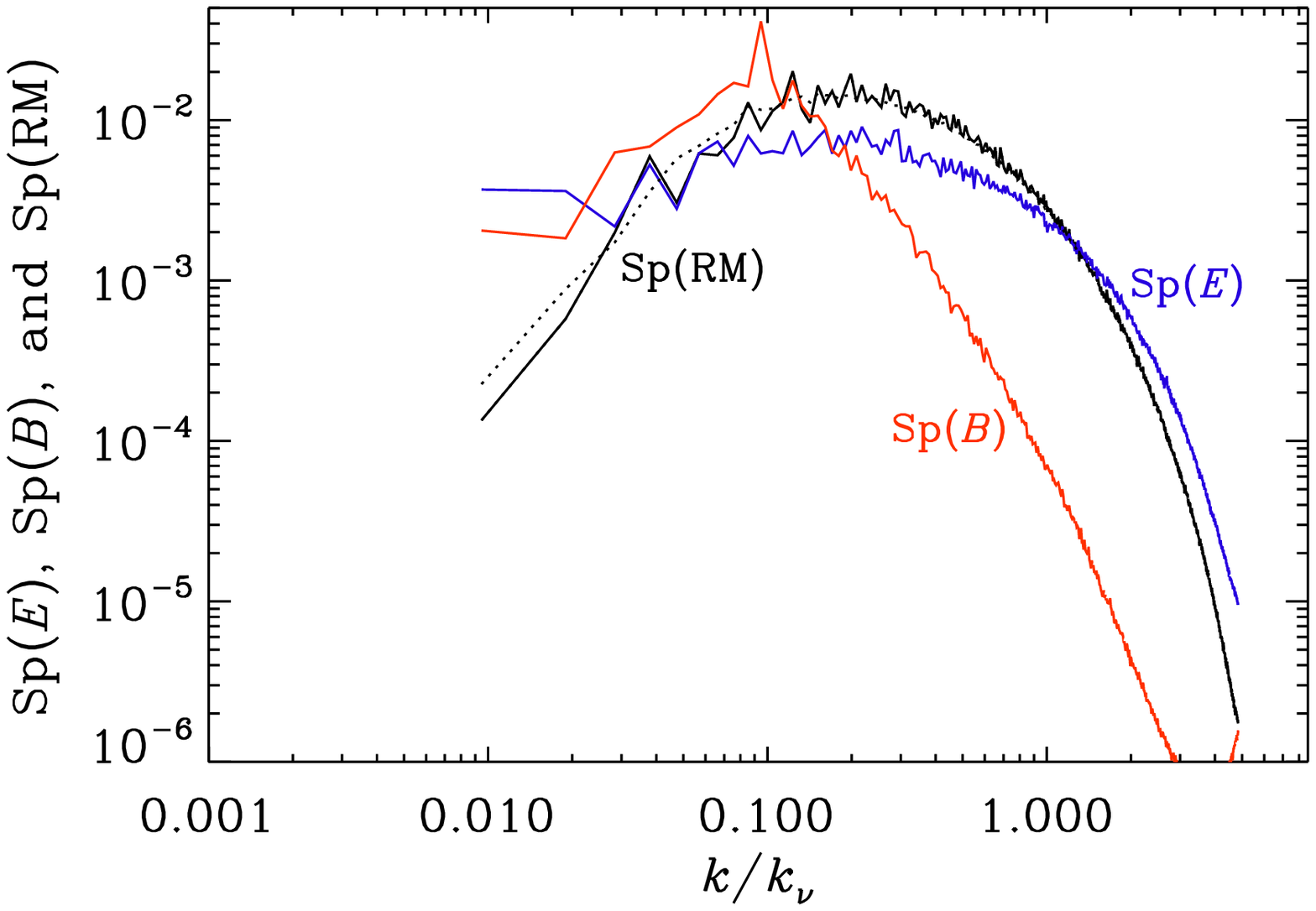}
\end{center}\caption[]{
Same as \Fig{pdiag_spec}, but for Run~C during the kinematic stage.
}\label{pdiag_spec_D1024_Pm10a_kf10}\end{figure}

\begin{figure}\begin{center}
\includegraphics[width=\columnwidth]{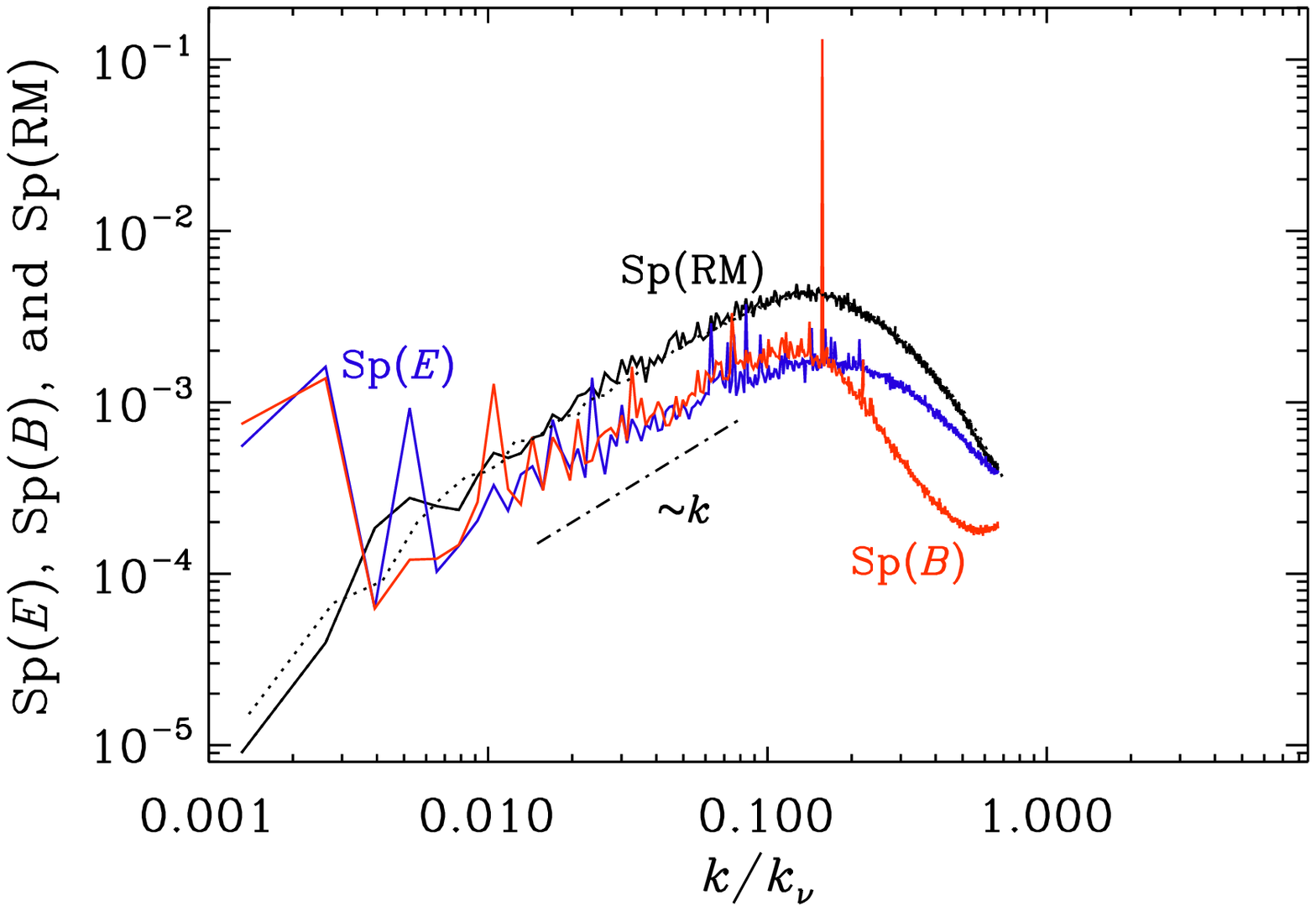}
\end{center}\caption[]{
Same as \Fig{pdiag_spec}, but for Run~A during the saturated stage.
}\label{pdiag_spec_D1024_Pm1e_kf120_sat}\end{figure}

\begin{figure}\begin{center}
\includegraphics[width=\columnwidth]{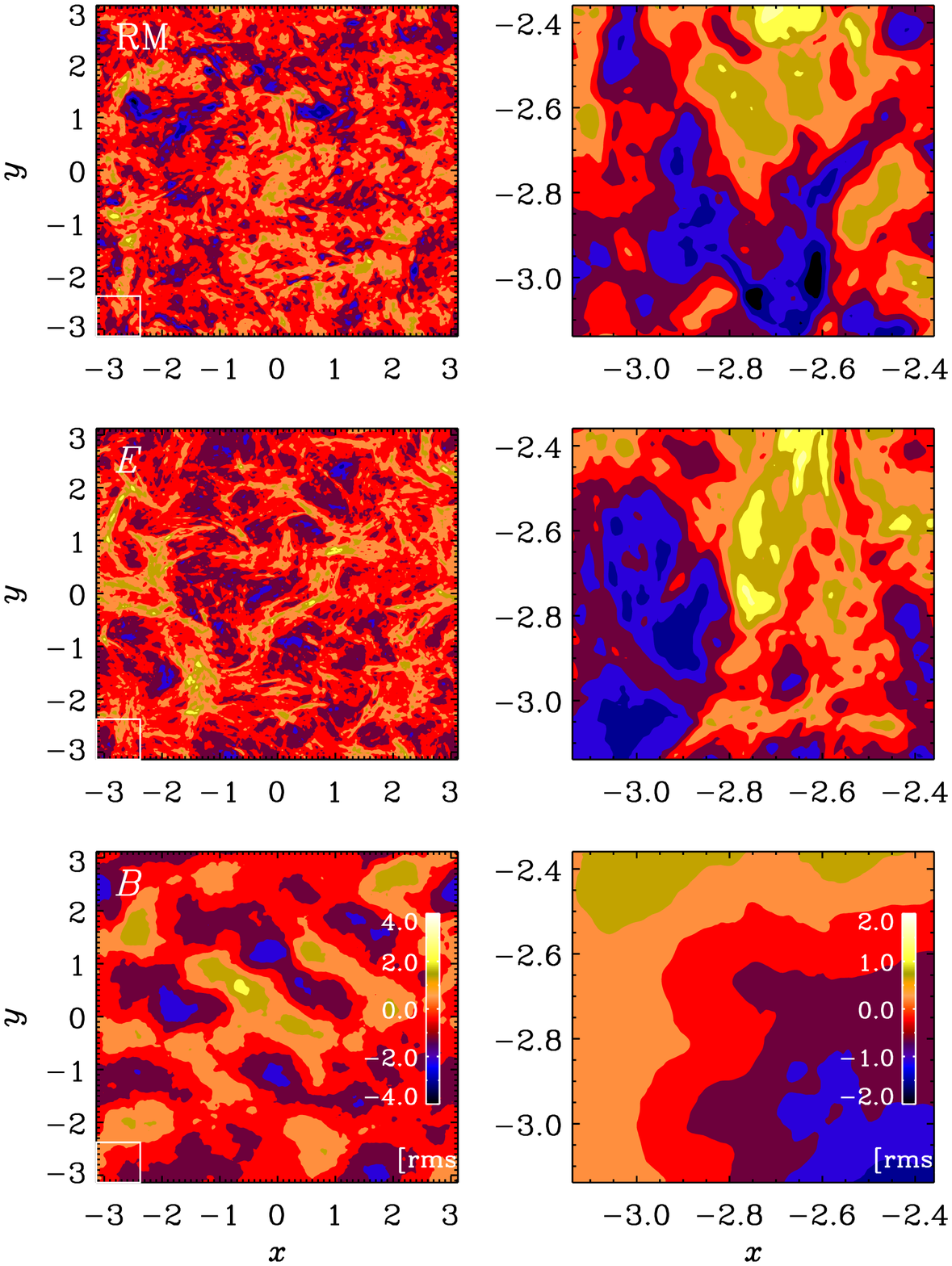}
\end{center}\caption[]{
Same as \Fig{pdiag_planes}, but for Run~E during the saturated stage.
}\label{pdiag_planes_D1024_Pm1c_kf4_sat}\end{figure}

\begin{figure}\begin{center}
\includegraphics[width=\columnwidth]{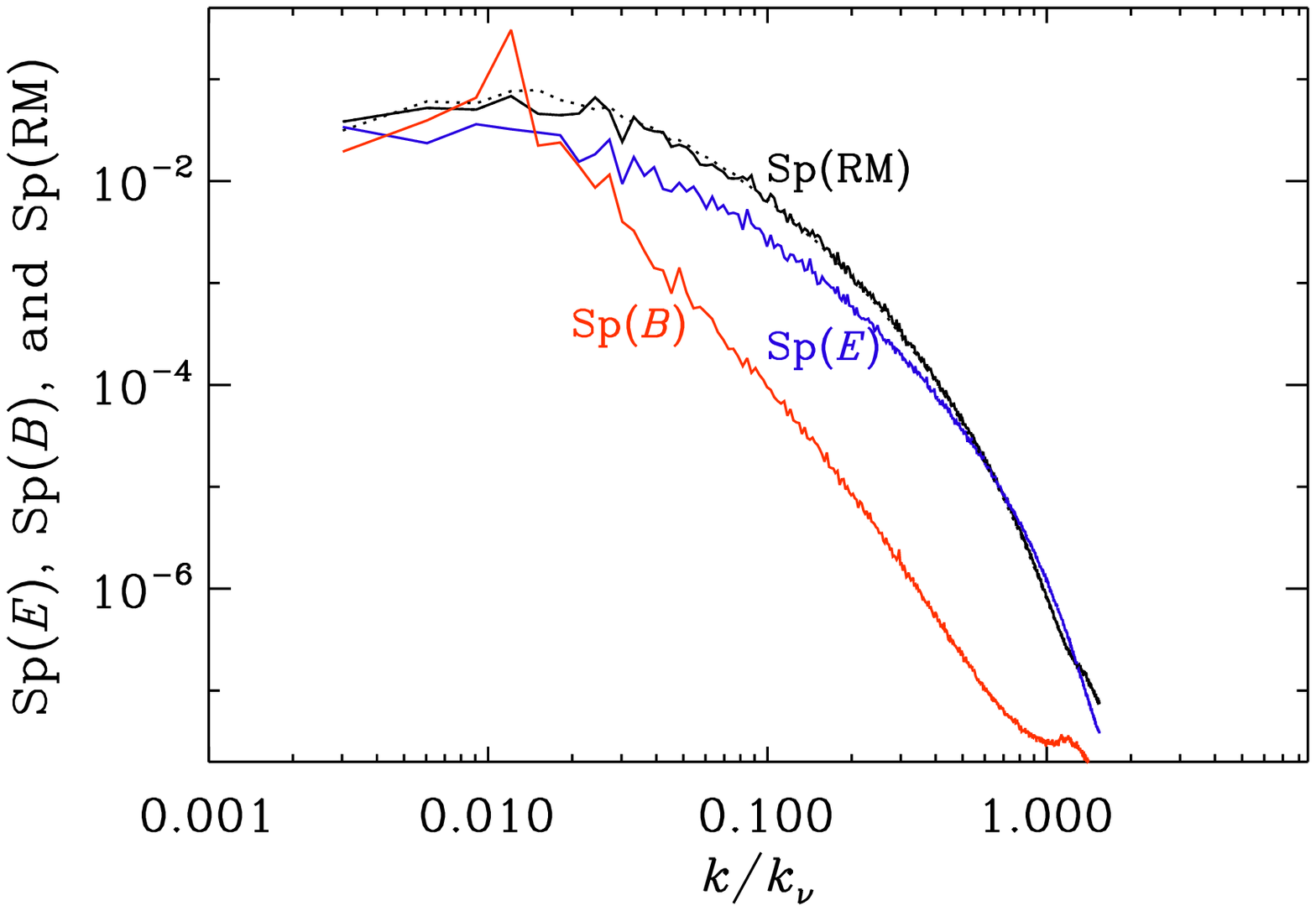}
\end{center}\caption[]{
Same as \Fig{pdiag_spec}, but for Run~E during the saturated stage.
}\label{pdiag_spec_D1024_Pm1c_kf4_sat}\end{figure}

In \Sec{Diagnostics}, we presented diagnostic images and spectra
for Runs~E, A, and D.
Here we also present spectra for Runs~B and C; see
\Figs{pdiag_spec_D1024_Pm1d_kf30}{pdiag_spec_D1024_Pm10a_kf10}.
Between Runs~B and C, we see a gradual increase in the elongated
structures in $E$, which is typical for all runs with $\Pm=1$.
These are all for the kinematic stage, but in this appendix we also
present results for the saturated stage of Runs~A and E; see
\Figss{pdiag_spec_D1024_Pm1e_kf120_sat}{pdiag_spec_D1024_Pm1c_kf4_sat}.
During saturation, the most remarkable change is seen in the
$B$ polarization of Run~A, which consists of stripes that are
inclined by $45\degr$.
This is caused by a systematic dominance of vertical field components
in the saturated state, which causes the formation of stacked clover leaf
patches in the $B$ polarization, as was demonstrated previously in the
appendix of \cite{BF20}.



\bsp	
\label{lastpage}
\end{document}